\documentclass{aa}  

\usepackage{graphicx}
\usepackage{txfonts}
\usepackage{xcolor}
\usepackage{CJKutf8}

\usepackage{array}
\newcolumntype{C}{>{\centering\arraybackslash}m{0.07cm}}
\usepackage{placeins}
\usepackage{hyperref}
\defcitealias{Izumi_2023}{I23}
\defcitealias{Tristram_2022}{T22}

\authorrunning{W. M. Goesaert et al.}
\begin{document} 
\begin{CJK*}{UTF8}{gbsn}

\title{Torus feeding and outflow launching in the active nucleus of the Circinus galaxy}
   \author{Wout M. Goesaert
          \inst{1}\fnmsep\inst{2}
          \and
          Konrad R. W. Tristram\inst{2}
          \and
          C. M. Violette Impellizzeri\inst{1,3}
          \and
          Alexander P. S. Hygate\inst{1}
          \and
          Sophie Venselaar\inst{1,4}
          \and
          Junzhi Wang (王均智)\inst{5}
          \and
          Zhiyu Zhang (张智昱)\inst{6,7}  
          }

   \institute{
            Leiden Observatory, University of Leiden, PO Box 9513, 2300 RA Leiden, The Netherlands
        \and
            European Southern Observatory, Alonso de Córdova 3107, Vitacura, Santiago, Chile
        \and
            Institute ASTRON, Netherlands Institute for Radio Astronomy, NL-7991 PD Dwingeloo, The Netherlands
        \and
            Instituto de Estudios Astrof\'{i}sicos, Universidad Diego Portales, Av. Ej\'{i}rcito Libertador 441, Santiago, Chile
        \and
            School of Astronomy and Space Science, Nanjing University, 163 Xianlin Avenue, Nanjing, 210023, People's Republic of China
        \and
            School of Astronomy and Space Science, Nanjing University, Nanjing 210093, China
        \and
            Key Laboratory of Modern Astronomy and Astrophysics (Nanjing University), Ministry of Education, Nanjing 210093, China
        }

   \date{Received May 20, 2025; accepted September 23, 2025}

  \abstract
   {Most active galactic nuclei (AGN) are believed to be surrounded by a dusty molecular torus on the parsec scale, which is often embedded within a larger circumnuclear disk (CND). AGN are fuelled by the inward transport of material through these structures and can launch multi-phase outflows that influence the host galaxy through AGN feedback.}
   {We use the Circinus Galaxy as a nearby laboratory to investigate the physical mechanisms responsible for feeding the torus and launching a multi-phase outflow in this Seyfert-type AGN, as these mechanisms remain poorly understood.}
   {We analysed observations from the Atacama Large Millimeter/submillimeter Array of the Circinus nucleus at angular resolutions down to 13 mas ($0.25\,\mathrm{pc}$). We traced dust and the ionised outflow using $86-665\,\mathrm{GHz}$ continuum emission and studied the morphology and kinematics of the molecular gas using the HCO$^+$(3-2), HCO$^+$(4-3), HCN(3-2), HCN(4-3) and CO(3-2) transitions.}
   {We find that the Circinus CND hosts molecular and dusty spiral arms, two of which connect directly to the torus. We detect inward mass transport along these structures and argue that the non-axisymmetric potential generated by these arms is the mechanism responsible for fuelling the torus. We estimate a feeding rate of $0.3-7.5\,\mathrm{M}_{\odot}\mathrm{yr}^{-1}$, implying that over 88\% of the inflowing material is expelled in a multi-phase outflow before reaching the accretion disk. The inferred feeding time scale ($t_{\mathrm{feed,\,torus}}=120\,\mathrm{kyr}-2.7\,\mathrm{Myr}$) suggests that variability in AGN activity may be driven by changes in torus feeding. On parsec scales, the ionised outflow is traced by optically thin free-free emission. The outflow is stratified, with a slightly wider opening angle in the molecular phase than in the dusty and ionised components. The ionised outflow is launched or collimated by a warped accretion disk at a radius of $r\sim0.16\,\mathrm{pc}$, and its geometry requires an anisotropic launching mechanism.}
   {}


   \maketitle
\end{CJK*}

\section{Introduction}
The Circinus galaxy hosts the nearest Seyfert~2 active galactic nucleus (AGN), located at a distance of $\sim 4\,\mathrm{Mpc}$ (1" $\sim 20\,\mathrm{pc}$). Active galactic nuclei are powered by the release of gravitational energy from material accreting onto a supermassive black hole (SMBH). This energy is released in the form of X-rays, UV radiation, and multi-phase outflows, including relativistic jets and molecular, dusty, and ionised winds. Through interactions with the surrounding medium, this energy output can profoundly influence the evolution of the host galaxy in a process known as AGN feedback (see, e.g., \citealt{Harrison_2024} and references therein). Understanding the physical mechanisms that fuel and regulate AGN activity is therefore critical for constraining models of coevolution between SMBHs and their host galaxies. Nearby AGN, such as in the Circinus galaxy, offer a unique laboratory for investigating these mechanisms in detail.

In the inner regions of Circinus, an accretion disk surrounding the SMBH can be traced accurately by H$_2$O megamasers \citep{Greenhill_2003}, revealing a significant warp at sub-pc scales and a SMBH mass of $M_{\bullet}=1.7\times10^6\,\mathrm{M}_{\odot}$. Combined with a bolometric luminosity of $L_{\mathrm{bol}} = 4\times 10^{43}\,\mathrm{erg\,s}^{-1}$ \citep{Moorwood_1996}, this implies an Eddington ratio of $\sim20\%$. Outside of this accretion disk, AGN are believed to contain a cold molecular dusty torus that surrounds it and can hide the high-energy radiation that is emitted by the accretion disk \citep{Krolik_1986, Antonucci_1993, Urry_1995}. Due to their small sizes, AGN tori can only be resolved at the highest angular resolutions, such as with the Atacama Large Millimeter/submillimeter Array (ALMA). Such observations yield information about their geometry, chemistry, and dynamics \citep{Burillo_2016, Izumi_2018, Combes_2019, Impellizzeri_2019}.

Because Circinus is a type~2 Seyfert \citep{Oliva_1994}, we observe this torus edge-on (\citealt[][hereafter I23]{Izumi_2023}; \citealt{Baba_2024}), which means that the broad line region is obscured from our view. As a result, we observe radiation that has been reprocessed by the structures surrounding the hot inner region of the AGN in the infrared (IR) and submillimetre regime \citep{Storchi_1992}. In the nucleus, the dominant mid-IR emitting structure is the dusty biconical outflow, which is perpendicular to the maser disk and has a size of $\sim2\,\mathrm{pc}$ \citep{Tristram_2007, Isbell_2022}. The ionised outflow, which is launched from the inner parsec region, is observed as a one-sided outflow cone in [O\,{\sc iii}] out to $\sim400\,\mathrm{pc}$ and as a two-sided plume in the radio out to $\sim1\,\mathrm{kpc}$ \citep{Marconi_1994, Elmouttie_1998}. Surrounding these AGN tori, we often find a CND that consists of cold molecular gas and dust. In the Circinus galaxy, the CND directly surrounds the torus. The torus and CND act as mass reservoirs from which AGN are fuelled \citep{Combes_2021b}.

High-resolution ALMA observations pave the way towards resolving some long-standing questions. From statistical studies of the obscuration fractions in AGN, we know, for example, that tori must have scale heights of $h/r\sim1$. However, the mechanism that keeps these tori geometrically thick remains unknown. The proposed mechanisms include supernova explosions within the torus \citep{Wada_2002}, radiation pressure from the thin disk \citep{Pier_1992}, magnetically driven outflows that sweep up obscuring material \citep{Emmering_1992} and the radiation-driven fountain model in which turbulence is injected into the torus by the backflow of the molecular outflow \citep{Wada_2012}. Other important open questions include the feeding mechanism that transports material from the CND down to the torus \citep{Combes_2021} and the launching mechanism of the ionised, dusty and molecular outflows, which could, for example, be driven by radiation pressure \citep{Dorodnitsyn_2011, Chan_2016, Honig_2017} or by magnetocentrifugal forces \citep{Vollmer_2018}. The aim of this work is to use the Circinus AGN as a laboratory to study torus feeding and outflow launching in a prototypical Seyfert~2 type AGN.

\section{Observations and data reduction}\label{sec:observations_data_reduc}
\subsection{Observations}
We obtained data from ALMA band 6 (259\,GHz, 1.16\,mm) and band 7 (350\,GHz, 0.86\,mm) of the Circinus nuclear region (project code 2018.1.00581.S, PI: K. Tristram). In band~6, short baseline (beam $\sim2\,\mathrm{pc}$) observations -- referred to as "B6SB" -- were obtained in addition to long baseline data (B6LB, beam $\sim0.4\,\mathrm{pc}$). In band 6, we targeted the HCO$^+$(3-2) and HCN(3-2) transitions, while in band~7, we targeted HCO$^+$(4-3), HCN(4-3) and CO(3-2). In addition, we use auxiliary data sets in bands~3 (86\,GHz \& 99\,GHz), band~4 (135\,GHz \& 147\,GHz), band~5 (178\,GHz) and band~9 (665\,GHz) (project codes 2017.1.00575.S,  PI: J. Wang and 2019.1.00013.S, PI: V. Impellizzeri) to investigate the spectral energy distribution (SED). We list observational details in Tables \ref{table:measurement_sets_primary} and \ref{table:measurement_sets_auxiliary} for the primary and auxiliary datasets.

\begin{table}[ht]
    \centering
    \caption{Observation and data reduction details of the primary datasets}            

    \begin{tabular}{c | c c c} 
        \hline
        \hline
         & B6SB & B6LB & B7\\
        \hline
        $\nu_{\mathrm{central}}$ [GHz] & 258.821 & 258.821 & 349.556\\
        $\lambda_{\mathrm{central}}$ [mm] & 1.16 & 1.16 & 0.86\\
        Beam [mas$^2$] & $127\times102$ & $21.5\times18.7$ & $49.0\times28.9$\\
        Beam [pc$^2$] & $2.5\times2.0$ & $0.4\times0.4$ & $1.0\times0.6$\\
        P.A. [$^{\circ}$] & -30.9 & 26.7 & 18.0\\
        MRS ["] & 1.616 & 0.375 & 0.595\\
        Int. time ['] & 37 & 90 & 93\\
        Sens. [$\mu$Jy/bm] & 22.0 & 13.0 & 40.9\\
        Obs. date & 15 Aug 2019 & 6 Jun 2019 & 13 Jul 2019\\
        Transitions & HCO$^+$(3-2) & HCO$^+$(3-2) & HCO$^+$(4-3)\\
        & HCN(3-2) & HCN(3-2) & HCN(4-3)\\
        & & & CO(3-2)\\
        Pipeline & 42254M & 42254M & 2021.2.0.128\\
        Self-cal iter.: & & & \\
        phase & 7 & 4 \& 6 & 5 \& 2\\
        amplitude & 1 & 1 \& 1 & 0 \& 0\\
        \hline
    \end{tabular}
    \tablefoot{The cited beam sizes represent those of the respective continuum maps imaged using Briggs weighting (robust parameter = 0.5). The maximal recoverable scales (MRS) were derived using Eq. (3.27) in \citet{ALMA_handbook}. The sensitivity is defined as the background rms noise measured in an emission-free region of the continuum map. Self-calibration iteration numbers separated by "\&" represent separate measurement sets that together make up one dataset. See Table \ref{table:measurement_sets_auxiliary} for the details on the auxiliary datasets in bands 3, 4, 5 and 9.} \label{table:measurement_sets_primary}
\end{table}

\subsection{Data reduction}\label{sec:data_reduc}
We performed the data reduction using the Common Astronomy Software Applications (CASA, \citealt{CASA}) to produce continuum and molecular emission maps. For band~6 and 7, we applied standard pipeline calibration (version listed in Table \ref{table:measurement_sets_primary}) using J1424-6807 as the bandpass calibrator and J1427-4206 as the phase and amplitude calibrator. We then applied several rounds of phase self-calibration on the continuum followed by one amplitude calibration -- except for the band~7 data due to an insufficient signal-to-noise ratio (S/N). We found that self-calibration increases the S/N by a factor of 9.9, 1.35, and 1.17 in the B6SB, B6LB, and B7 datasets, respectively. After self-calibration, we subtracted the continuum emission using "\texttt{fitorder=0}" in the "\texttt{imcontsub}" casa task. Since the B6LB and B7 datasets were split into two separate observing epochs, we combined these epochs into a single measurement set using the "\texttt{concat}" task in CASA. We also explored the possibility of combining B6SB with B6LB, but decided not to do this due to an offset of $\sim30\,\mathrm{ mas}$ between the peak of the continuum of B6SB and those of B6LB and B7 (see Sect. \ref{sec:Astrometry}).

We then imaged the datasets using the "\texttt{tclean}" casatask with auto-masking\footnote{We use \texttt{sidelobethreshold=3.0}, \texttt{noisethreshold=5.0}, \texttt{minbeamfrac=0.3} and \texttt{lownoisethreshold=2.0}. For continuum imaging, we set \texttt{negativethreshold=0.0} and used the Hogbom algorithm. For line imaging, we instead used \texttt{negativethreshold=7.0} and multi-scale cleaning.}, velocity bins of $10.2\,\mathrm{km\,s^{-1}}$ for B6SB and $7.6\,\mathrm{km\,s^{-1}}$ for B6LB and a cell size of $3\,\mathrm{mas}$. We used Briggs weighting with robust parameters of 0.5 and 2.0 for the continuum and molecular emission maps, respectively, to optimise S/N and resolution. In addition, we cleaned the B6LB continuum using a robust parameter of -1.0 to image the continuum morphology at a beam size of only $14.3\times12.2\,\mathrm{mas}^2$. After imaging, we applied primary beam correction to all maps, which boosted the flux density by less than one per cent at the edge of the image. Lastly, we created moment-zero maps from the molecular line data cubes using the software CARTA \citep{CARTA}.

\section{Results}\label{sec:results}

\begin{figure}[h]
    \centering
    \includegraphics[width=\linewidth]{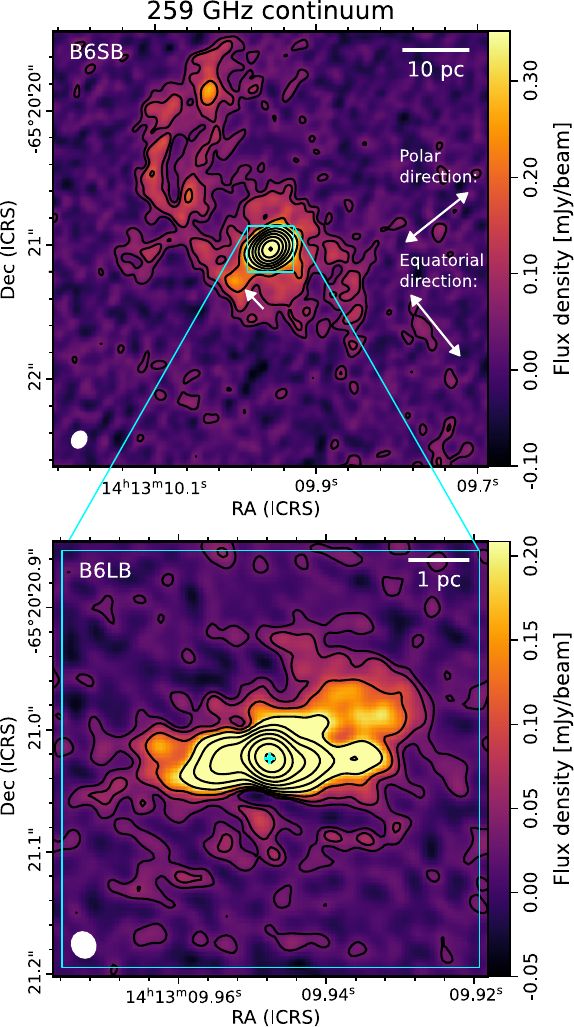}
    \caption{Band 6 short (top) and long (bottom) baseline continuum maps at 259 GHz. Contour levels are drawn at $[2, 4, 8, 16, $ $32, 64, 128, 256, 512, 1024]\,\times\,$rms, where the background rms equals $0.022\,\mathrm{mJy/beam}$ in the B6SB map and $0.013\,\mathrm{mJy/beam}$ in the B6LB map. The synthesised beam sizes are drawn in the bottom left as filled ellipses, and the cyan cross indicates the AGN position (as defined in Sect. \ref{sec:Astrometry}). The phoenix feature is indicated with a white arrow. The polar and equatorial directions mentioned throughout the text are shown in the upper panel.}
    \label{fig:cont_morphology}
\end{figure}

\subsection{Continuum emission}\label{sec:results_continuum}
We present the continuum maps of B6LB and B6SB at $259\,\mathrm{GHz}$ in Fig. \ref{fig:cont_morphology}. We do not show the B7 continuum here, since its morphology is very similar to that of B6. An overview of all continuum maps between 86\,GHz and 665\,GHz is presented in Fig.~\ref{fig:cont_SED}. The short baseline data of B6 show an unresolved core that is immediately surrounded by a $\sim10\,\mathrm{pc}$ long elongated component with a position angle (PA) of $\sim 130^{\circ}$. This direction matches the PA of the dusty outflow on the pc scale \citep{Stalevski_2019} and the one-sided ionised outflow cone that extends to $\sim1\,\mathrm{kpc}$ \citep{Marconi_1994, Kakkad_2023}. We refer to this as the polar direction, as indicated in Fig.~\ref{fig:cont_morphology}. Toward the south-east, we observe an elongated component along this polar axis that extends to a projected distance of $\sim8\,\mathrm{pc}$ from the AGN. We refer to this structure as the "Phoenix"\footnote{We name this structure the Phoenix since this protrusion ends in a flux enhancement that resembles a head. At this head, the structure splits into two "wings".} (white arrow in Fig. \ref{fig:cont_morphology}).
The nuclear core is unresolved with short baselines and shows a flux density of 23.3\,mJy but is resolved by the long baseline data in band 6 into a central peak of 14.6\,mJy with extended polar structure. Following the levels of the continuum contours inward from $\sim10\,\mathrm{pc}$ to the unresolved $0.4\,\mathrm{pc}$ core, we observe a clockwise rotation. The contours rotate from the polar direction ($\mathrm{PA}\sim 130^{\circ}$) in B6SB to a $\sim3\,\mathrm{pc}$ long horizontal bar ($\mathrm{PA}\sim 92^{\circ}$) in B6LB and eventually to the equatorial direction ($\mathrm{PA}\sim 45^{\circ}$, as indicated in Fig. \ref{fig:cont_morphology}) at the parsec scale.

On an arc second scale, we observe the CND which traces an S shape and closely matches the morphology of the molecular CND. To the north-east, we detect a spiral arm that extends out to $\sim30\,\mathrm{pc}$ and connects to the CND. These extended structures are visible in bands~6, 7, and 9, but are not detected at lower frequencies despite similar sensitivities (see Fig. \ref{fig:cont_SED}).

\subsection{Molecular line emission}\label{sec:results_molecular}
\begin{figure}[h]
    \centering
    \includegraphics[width=\linewidth]{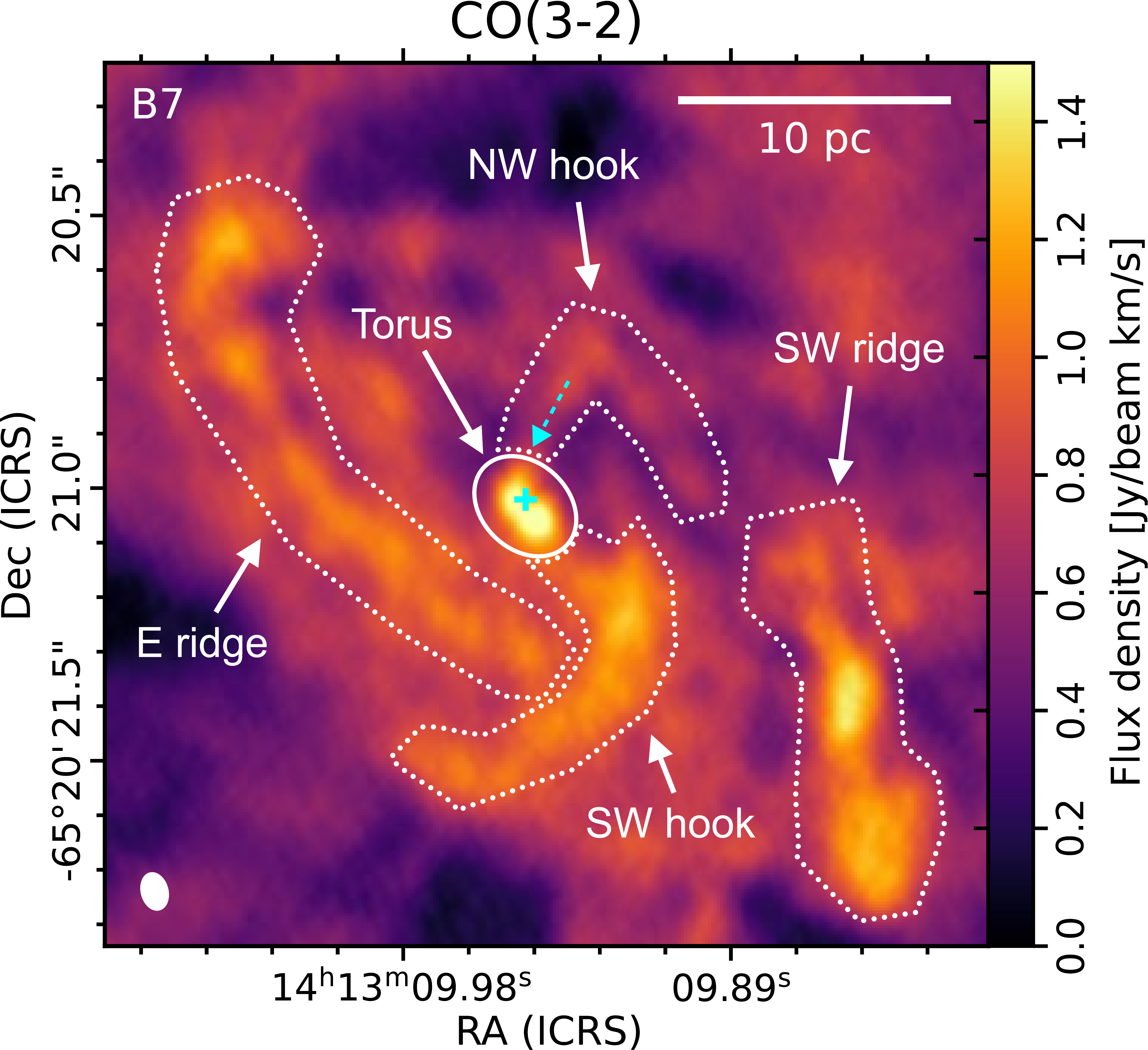}
    \caption{Spiral structures in the circumnuclear disk (as informed by the morphology in Figs. \ref{fig:mol_morphology} and \ref{fig:mol_morphology_B7}) overlaid on the band 7 CO(3-2) ($\nu_{\mathrm{rest}}=345.8\,\mathrm{GHz}$) map. The dashed cyan arrow represents the proposed inflow through the NW hook that feeds the torus as discussed in Sect. \ref{sec:inflow}. The synthesised beam size is drawn in the bottom left as a filled ellipse, and the cyan cross indicates the AGN position.}
    \label{fig:CO_annotated}
\end{figure}

\begin{figure*}[ht]
    \centering
    \includegraphics[width=\linewidth]{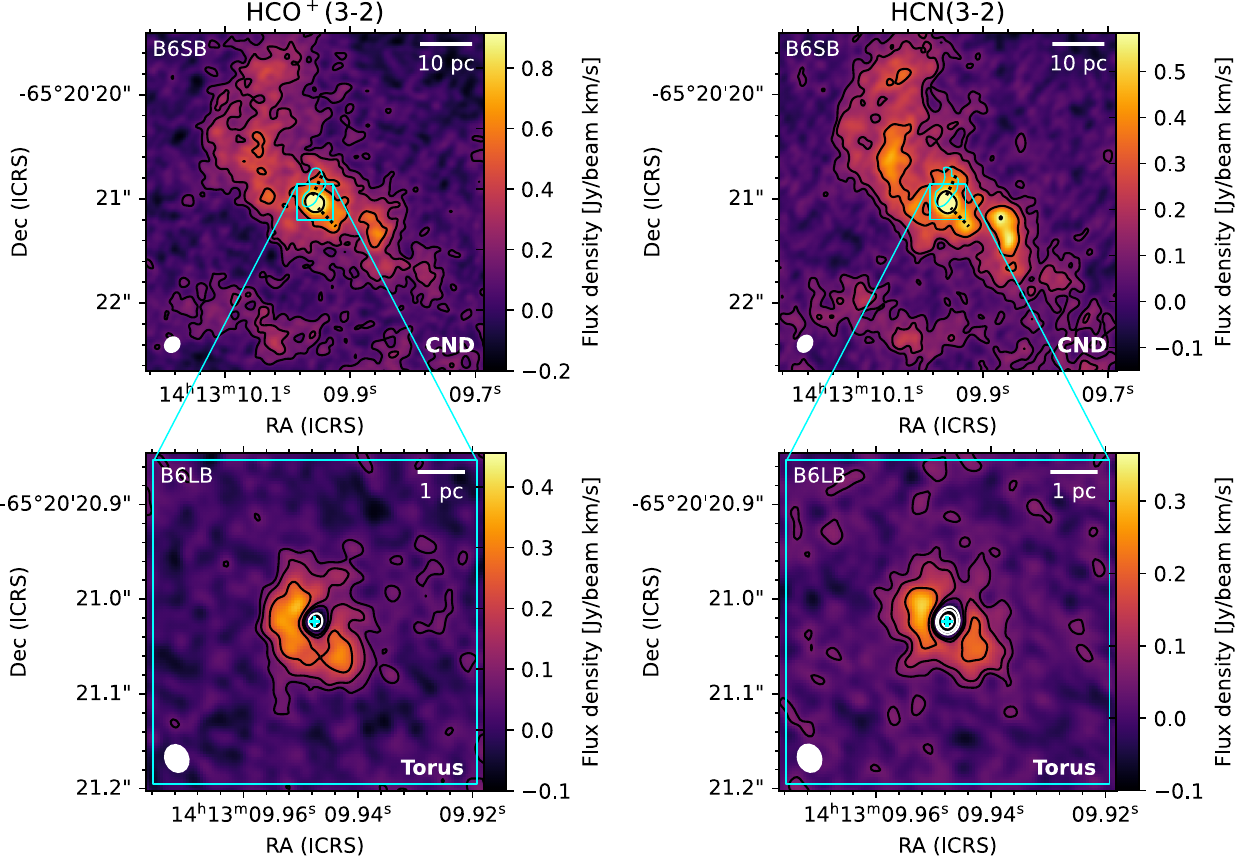}
    \caption{Band 6 HCO$^+$(3-2) ($\nu_{\mathrm{rest}}=267.6\,\mathrm{GHz}$) and HCN(3-2) ($\nu_{\mathrm{rest}}=265.9\,\mathrm{GHz}$) short (top, covering the CND) and long (bottom, showing the resolved torus) baseline moment zero maps. Contour levels are drawn at $[-16, -8, -4, 2, 4, 8, 16]\,\times\,$rms, where the background rms equals 36, 23, 57 and $28\,\mathrm{mJy/beam\,km/s}$ in the HCO$^+$(3-2) B6SB and B6LB and HCN(3-2) B6SB and B6LB maps, respectively. White contours represent negative flux density values and indicate the position of an absorption hole that coincides with the continuum peak. The L-shape that contains the NW and SW hooks as discussed in the text is drawn as black dashed lines in the top panels. Cyan contours in the top panels indicate the position of the blue-shifted emission feature discussed in Sect. \ref{sec:inflow} and represent the $-30\,\mathrm{km\,s^{-1}}$ level in the right panels of Fig. \ref{fig:BBAROLO_velfield}. The synthesised beam sizes are drawn in the bottom left as filled ellipses, and the cyan cross indicates the AGN position.}
    \label{fig:mol_morphology}
\end{figure*}

We present the CO(3-2) molecular map zoomed in on the CND in Fig.~\ref{fig:CO_annotated} and show the HCO$^+$(3-2) and HCN(3-2) maps in Fig.~\ref{fig:mol_morphology}. In addition, we present the HCO$^+$(4-3), HCN(4-3) and CO(3-2) maps in Fig. \ref{fig:mol_morphology_B7}. The HCO$^+$(3-2) and HCN(3-2) transitions trace dense molecular gas ($n_{\mathrm{crit,\,HCO^+(3-2)}}= 1.0\times10^6\,\mathrm{cm}^{-3}$ and $n_{\mathrm{crit,\,HCN(3-2)}}= 5.7\times10^6\,\mathrm{cm}^{-3}$ at $T_{\mathrm{kin}}=50\,\mathrm{K}$, \citealt{Shirley_2015}), while CO(3-2) traces slightly lower density molecular gas ($n_{\mathrm{crit,\,CO(3-2)}}\approx 1.7\times10^4\,\mathrm{cm}^{-3}$ at $T_{\mathrm{kin}}=50\,\mathrm{K}$)\footnote{Based on the Einstein A coefficients and collisional rates from the Leiden Atomic and Molecular Database (LAMBDA;
\citealt{LAMDA_2005, Lamda_CO_collrates}).}.

On a scale of $\gtrsim 10\,\mathrm{pc}$, the dense molecular gas traces a morphology that is very similar to that of the B6SB continuum, except for the Phoenix, which is absent from the molecular maps. At the centre of the CND, we find the molecular torus, which is viewed almost edge-on (inclination of $i\sim 85^{\circ}$ as determined by \citealt{Baba_2024} and \citetalias{Izumi_2023}). We measure the projected size of the torus to be approximately $120\times90\,\mathrm{mas}^2$ or $2.4\times1.8\,\mathrm{pc}^2$ from the HCO$^+$ disk, which appears slightly more extended than the HCN disk. The emission from this edge-on torus resembles that of an oblate disk with an "absorption hole" at the centre that coincides with the continuum peak, as also described by \citepalias{Izumi_2023}. In the lower-resolution HCO$^+$(4-3), HCN(4-3) and CO(3-2) maps, we detect a local reduction in surface brightness rather than negative flux density. The emission from the torus disk appears inhomogeneous, being dimmer to the north-west than to the south-east (in all maps, but most apparently in HCO$^+$(3-2)). Both HCN(3-2) and  HCO$^+$(3-2) show an emission peak directly to the north-west and decreased emission to the south and north-west.

\subsubsection{Spiral structure in the CND}\label{sec:spiral_structure}
The molecular tracers in this study reveal a spiral structure in the CND, which we divide into four distinct structures (Fig. \ref{fig:CO_annotated}). South-west of the torus, we observe the "SW ridge", as also described in \citet[][hereafter T22]{Tristram_2022}. In the B6LB and B7 maps, this SW ridge is resolved into three $\sim4\,\mathrm{pc}$ clumps. We only observed the two northernmost clumps in the high-density gas tracers HCN and HCO$^+$, implying that their density is higher than that of the southern clump. To the east, we observe a 20\,pc long ridge-like substructure within the CND, which we refer to as the "eastern ridge".

Within the central 10\,pc surrounding the torus, we observe an L shape in the HCO$^+$(3-2) and HCN(3-2) maps with one side extending to the north of the torus and one side extending to the south-west (black dashed lines in Fig.~\ref{fig:mol_morphology}). In the CO(3-2), HCO$^+$(4-3) and HCN(4-3) maps, this structure is resolved into two "hook-like" structures that each consist of an inner straight bar and an outer spiral arm (see Fig.~\ref{fig:mol_morphology_B7}). We refer to these as the "NW hook" and the "SW hook" (as indicated in Fig. \ref{fig:CO_annotated}). Together, these ridges and hooks form a spiral structure in the CND, not too different in morphology from the spiral structures expected at these scales from hydrodynamic simulations \citep{Wada_2012, Wada_2016}.

\subsection{Astrometry}\label{sec:Astrometry}
We find that the location of the peaks in the continuum maps of B6LB (259\,GHz) and B7 (350\,GHz) are consistent within $3\,\mathrm{mas}$, the size of the imaging cell. Since these are independent measurements, we thus conclude that the continuum peak is located at $\alpha=14^{\mathrm{h}}13^{\mathrm{m}}09.9473\pm0.0005^{\mathrm{s}}$, $\delta=-65^{\circ}20'21.023\pm0.003"$ (ICRS). This uncertainty of $3\,\mathrm{mas}$, which is $\approx17\%$ of the B6LB beam size, represents the maximum absolute astrometric precision that can be reached with ALMA using the standard calibration procedure (see Sect. A.9.5 in the ALMA cycle 6 Proposer's Guide; \citealt{Andreani_2018}). This continuum peak likely coincides with the AGN location, tracing the compact X-ray corona or the innermost part of the accretion disk (see Sect. \ref{sec:nuclear_free_free}), so we refer to this as the "AGN position" throughout this paper. There could be an offset between the kinematic centre (and thus the AGN position) and the continuum peak. However, in Sect.~\ref{sec:VLBI_astrometry}, we show that any such offset is likely smaller than $6\,\mathrm{mas}$. Our AGN position is consistent with \citetalias{Tristram_2022}, who concluded that the AGN is located at $\alpha=14^{\mathrm{h}}13^{\mathrm{m}}09.942\pm0.016^{\mathrm{s}}$, $\delta=-65^{\circ}20'21.03\pm0.10"$. We therefore also confirm the discrepancy already noted by \citetalias{Tristram_2022} with the astrometry from \citet{Greenhill_2003}, who found that the kinematic centre lies about $180\,\mathrm{mas}$ to the south of our continuum peak at $\alpha=14^{\mathrm{h}}13^{\mathrm{m}}09.95\pm0.02^{\mathrm{s}}$, $\delta=-65^{\circ}20'21.2\pm0.1"$.

For B6SB, we find that the continuum peak is located at $\alpha=14^{\mathrm{h}}13^{\mathrm{m}}09.942\pm0.0024^{\mathrm{s}}$, $\delta=-65^{\circ}20'21.030\pm0.015"$, which is offset $\sim30\,\mathrm{mas}$ ($\approx25\%$ of the B6SB beam size) from the location in the B6LB and B7 datasets. We consider the B6LB and B7 continuum peak positions to be more reliable due to their smaller beam size and consistent astrometry.

\section{Discussion}\label{sec:discussion}
\subsection{Continuum emission mechanisms}\label{sec:continuum_origin}

\begin{figure*}
    \sidecaption
    \includegraphics[width=12cm]{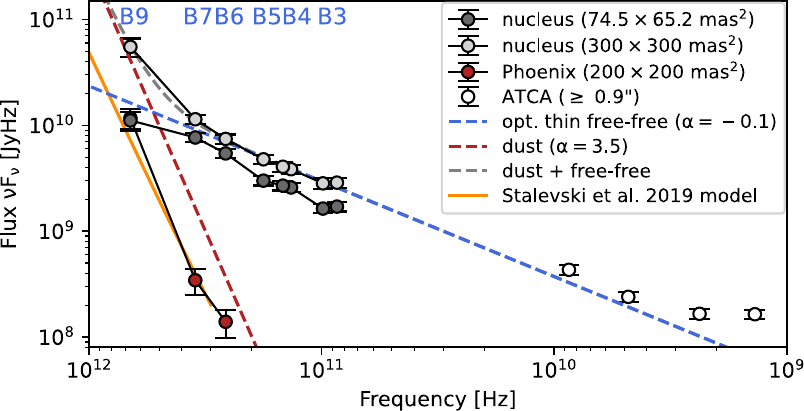}
    \caption{Spectral energy distribution of the Circinus AGN for the three apertures drawn in Fig. \ref{fig:FF_dust_model}. Also shown are four ATCA measurements at varying beam sizes \citep{Elmouttie_1998} in white. The dashed lines represent a power-law ($F_{\nu}\propto \nu^{\alpha}$) fit of the 300 mas nuclear aperture containing both a dust component with $\alpha=3.5$ and a free-free component with $\alpha=-0.1$. We also plot the nuclear dust model by \citet{Stalevski_2019} in orange.}
    \label{fig:SED}
\end{figure*}

\begin{figure*}
    \centering
    \includegraphics[width=0.96\linewidth]{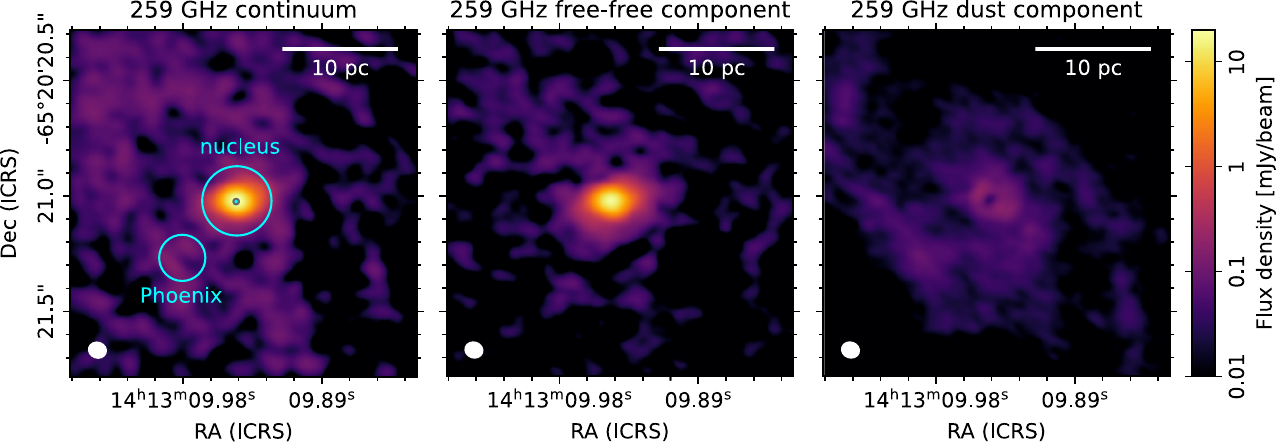}
    \caption{Band 6 (259 GHz) continuum decomposed into a free-free ($\alpha=-0.1$) and a dust ($\alpha=3.5$) component based on pixel-by-pixel SED-fitting. The common beam size between all maps used for these SED fits is drawn in the bottom left as a filled ellipse. In the left figure, we indicate the Phoenix and the two nuclear apertures used to obtain the spectral energy distribution shown in Fig. \ref{fig:SED}.}
    \label{fig:FF_dust_model}
\end{figure*}

To investigate the continuum emission mechanism, we constructed an SED between 86\,GHz (band~3) and 665\,GHz (band~9). For this, we first convolved all continuum maps listed in Tables \ref{table:measurement_sets_primary} and \ref{table:measurement_sets_auxiliary} to a common beam size of $74.5\times65.2\,\mathrm{mas}^2$ (see Fig. \ref{fig:cont_SED} for an overview of these maps). We then extracted SEDs in three regions: the continuum peak at the AGN position, a circular aperture with a diameter of 300\,mas also centred on the AGN, and a smaller circular aperture of 200\,mas centred on the continuum peak of the Phoenix structure (as indicated in Fig. \ref{fig:FF_dust_model}). We determined the errors of these flux measurements by adding the background rms from an empty region at the edge of each map to the absolute flux error. We also added in quadrature the absolute flux error of $10\,\%$ ($20\,\%$ for band~9), as stated in the ALMA technical handbook \citep{ALMA_handbook}. The resulting SED is presented in Fig. \ref{fig:SED} and the flux density measurements are listed in Table \ref{table:SED_data}. We also show in Fig.~\ref{fig:SED} lower-resolution nuclear ATCA measurements between 1.5\,GHz and 10\,GHz from \citet{Elmouttie_1998}, as there are no high-resolution radio observations available in the literature.

The Phoenix SED resembles a power law that is consistent with pure dust emission ($\alpha=[3-4]$ for $F_{\nu}\propto \nu^{\alpha}$; \citealt{Hildebrand_1983, Kawamuro_2022}). The nucleus shows a power law with a different slope that is consistent with optically thin free-free emission from ionised gas ($\alpha = -0.1$; \citealt{Draine_2011}). The 300\,mas nuclear aperture contains more flux than the central beam, indicating that this free-free component is resolved. We interpret the nuclear emission as free-free radiation rather than synchrotron radiation based on the SED slope and the brightness temperature, which is too low for synchrotron emission ($T_{\mathrm{B}}\leq1.3\times10^3\,\mathrm{K}$; see Sect.~\ref{sec:synchrotron_contribution} for a discussion on the possible synchrotron contribution from a nuclear jet or the X-ray corona). In B7 and B9, the 300\,mas nuclear SED shows a flux density excess compared to the free-free component. As discussed in \citetalias{Tristram_2022}, this excess is likely caused by thermal emission from cold dust in the equatorial plane surrounding the AGN. This excess cannot be caused by synchrotron radiation from the X-ray corona -- as observed in NGC~1068 \citep{Inoue_2020, Mutie_2025} -- because we observe no excess at the continuum peak.

To further investigate the morphology of the dust and ionised gas, we decomposed the emission into a dust and a free-free component. We did this for every pixel by performing a least-squares fit that combines two power-law components with $\alpha=-0.1$ and $\alpha=3.5$ for free-free and dust emission, respectively. We plot the equivalent flux density of each component at 259\,GHz in Fig.~\ref{fig:FF_dust_model}. We find that the free-free component traces a nuclear structure that is elongated in the polar direction. The dust component traces a spiral structure similar to that observed in the molecular disk, as well as dust surrounding the AGN in the equatorial direction. From this pixel-based SED decomposition, we find that at $259\,\mathrm{GHz}$, 98\% of the flux contained within the $4\times\mathrm{rms}$ contour in Figs.~\ref{fig:cont_morphology} and \ref{fig:outflow_morphology} is free-free emission from nuclear ionised gas.

The lower resolution ATCA measurements lie near the nuclear $\alpha=-0.1$ power law when extrapolated from the millimetre (mm) regime towards 1\,GHz (as also pointed out by \citetalias{Tristram_2022}). From the inferred electron densities and temperatures (Sect.~\ref{sec:nuclear_free_free}), we expect the nuclear free-free emission to become optically thick around $\sim3-19\,\mathrm{GHz}$ (from Eq. \ref{eq:FF_optdepth}). Therefore, ATCA measurements contain a significant contribution from nuclear free-free emission but must be dominated by a different emission component at the lowest frequencies ($<3\,\mathrm{GHz}$). This could, for example, be radiation from nuclear star formation (e.g. in the SW ridge; \citealt{Mueller_2006}), self-absorbed synchrotron radiation from a weak jet or diffuse synchrotron emission from cosmic-ray electrons \citep{Condon_1992, Mutie_2025}. Higher-resolution low-frequency observations are needed to constrain these scenarios.

\subsection{Nuclear free-free emission}\label{sec:nuclear_free_free}
On sub-pc to pc scales, nuclear free-free emission can originate from the compact hot X-ray corona, X-ray heated gas surrounding the accretion disk or from the ionised outflow \citep{Haardt_1991, Palit_2024, Vollmer_2018}. The resolved free-free emission on the pc scale likely traces the ionised outflow because (1) it is orientated along the polar axis, (2) its morphology resembles that of the dust outflow cone, and (3) it is co-spatial with the outflow maser spots (see Fig.~\ref{fig:outflow_morphology}). The unresolved core, on the other hand, may either be a compact point source (e.g. the X-ray corona) or a disk-like structure, as observed by \citet{Gallimore_2004} in NGC 1068 (component "S1").

The pc-scale outflow has a brightness temperature of $T_{\mathrm{B}}\sim1-40\,\mathrm{K}$ at $\nu=259\,\mathrm{GHz}$. At this distance from the AGN, we expect the electron temperature to be near the equilibrium photo-ionisation temperature \citep{Wada_2018}. We thus assume $T_{\mathrm{e}}=\left(0.8-3\right)\times10^4\,\mathrm{K}$, which implies an optical depth of $\tau=-\mathrm{ln}\left[1 - (T_{\mathrm{B}}/T_{\mathrm{e}})\right]=\left(0.33-50\right)\times10^{-4}$. Further assuming a homogeneous electron density $n_{\mathrm{e}}$ along an optical path of length $l=1\,\mathrm{pc}$, we can estimate $n_{\mathrm{e}}$ from \citep{Altenhoff_1960}:

\begin{equation}\label{eq:FF_optdepth}
    \begin{aligned}    \tau_{\mathrm{FF}}=3.278\times10^{-7}\left(T_{\mathrm{e}}/10^4\,\mathrm{K}\right)^{-1.35}\left(\nu/\mathrm{GHz}\right)^{-2.1}\,n_{\mathrm{e}}^2\,\left(l/\mathrm{pc}\right),
    \end{aligned}
\end{equation}

which yields $n_e=\left(6-46\right)\times10^3\,\mathrm{cm}^{-3}$. In the hydrodynamic model of \citet{Wada_2018}, ionised gas with $T_e \sim 10^4\,\mathrm{K}$ and $n_e \sim 10^4\,\mathrm{cm}^{-3}$ is predicted to exist in a wide-angle cone region surrounding the narrow-line region. This electron density is also comparable to the value derived by \citetalias{Izumi_2023} for the $\mathrm{H}36\alpha$-traced ionised outflow in the Circinus AGN ($n_e \sim 7400\,\mathrm{cm}^{-3}$). Assuming $T_{\mathrm{e}} = 10^4\,\mathrm{K}$ and $l = 2\,\mathrm{pc}$ as in \citetalias{Izumi_2023}, we find $n_e = (4.2$–$27)\times10^3\,\mathrm{cm}^{-3}$.

The unresolved core has a much higher antenna temperature\footnote{$T_{\mathrm{B}}$ represents the intrinsic surface brightness, while $T_{\mathrm{A}}$ is the measured surface brightness. In general, $T_{\mathrm{B}}\geq T_{\mathrm{A}}$ due to beam dilution.}, reaching $T_{\mathrm{A}}=1.3\times10^3\,\mathrm{K}$ at the highest B6LB resolution (middle panel of Fig. \ref{fig:outflow_morphology}). Such a high antenna temperature implies that $T_e$ is likely higher than the photo-ionisation temperature because at $T_e=10^4\,\mathrm{K}$, the free-free emission becomes optically thick for $\nu\leq101\,\mathrm{GHz}$, whereas the observed SED slope remains consistent with optically thin emission down to $\nu=86\,\mathrm{GHz}$. If, for example, we assume $T_e\leq2\times10^4\,\mathrm{K}$, then the implied optical depth at $\nu=86\,\mathrm{GHz}$ is $\tau\geq0.68$, which would reduce the flux density by $\geq30\%$ compared to the optically thin case\footnote{In the optically thin case, 60\% of the flux density in the $86\,\mathrm{GHz}$ common beam would correspond to the B6LB unresolved core and 40\% to the ionised outflow. The total attenuation thus equals $0.6\times\left(1-e^{-\tau}\right)$.}. No such reduction in flux density is detected at $\geq3\,\sigma$ significance, so this low-temperature scenario is disfavoured. At $r<0.25\,\mathrm{pc}$, strong X-ray heating is instead expected to drive the gas to the Compton equilibrium temperature \citep{Krolik_1986}. Assuming $T_{\mathrm{e}}\sim6\times10^6\,\mathrm{K}$ analogous to the "S1" region in NGC~1068 \citep{Gallimore_2004} and taking $l=0.25\,\mathrm{pc}$ (the beam size), Eq.~(\ref{eq:FF_optdepth}) yields a much higher electron density of $n_{\mathrm{e}}\sim1.3\times10^6\,\mathrm{cm}^{-3}$, which is similar to that found by \citet{Gallimore_2004} in NGC~1068 ($n_e\sim8\times10^5\,\mathrm{cm}^{-3}$).

We conclude that the pc-scale free-free emission and the extended $\mathrm{H}36\alpha$ emission trace the ionised outflow at similar density, while the unresolved core represents a distinct, higher-density and temperature component, akin to the "S1" region in NGC~1068 (X-ray heated gas in the corona or above the accretion disk).

\subsubsection{Ionised outflow}\label{sec:ionised_outflow}
\begin{figure}
    \centering
    \includegraphics[width=\linewidth]{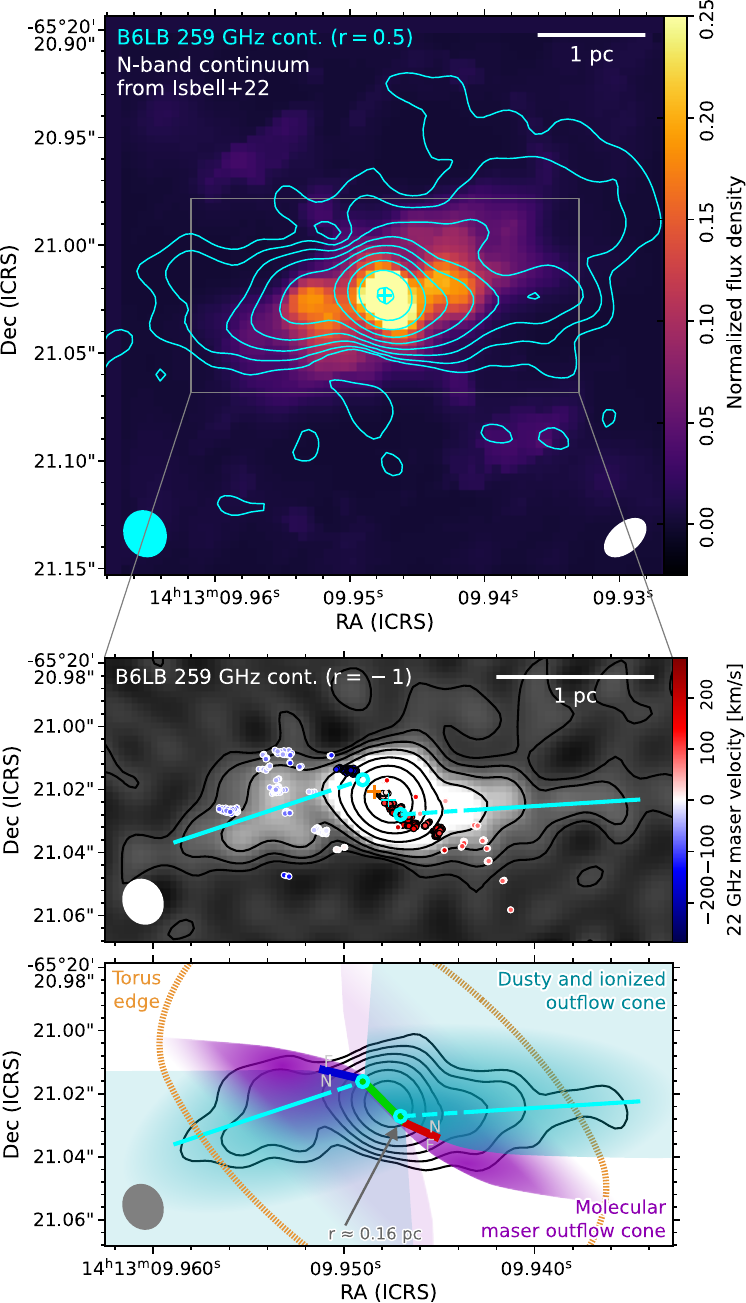}
    \caption{Ionised gas morphology (traced by the 259 GHz continuum) compared to the dusty outflow cone (traced by the mid-IR N-band continuum from \citealt{Isbell_2022}, top panel) and the accretion disk and molecular outflow (traced by the 22 GHz mega-masers from \citealt{McCallum_2009}, middle panel). The B6LB 259 GHz continuum (robust parameter $\,=0.5$) is drawn in the top panel as cyan contours at $[4, 8, 16, 32, 64, 128, 256, 512]\,\times\,$rms, with $\mathrm{rms}=0.013\,\mathrm{mJy/beam}$. The B6LB and N-band continuum synthesised beam sizes are shown on the bottom left and right, respectively. The middle panel shows the higher resolution (robust parameter $\,=-1$) B6LB continuum map with contour levels at $[2, 4, 8, 16, 32, 64, 128]\,\times\,$rms, with $\mathrm{rms}=0.048\,\mathrm{mJy/beam}$. Overlaid on the middle panel are the 22 GHz maser spots that trace the Keplerian disk (black edge) and the outflow (white edge). The maser spots have a global positional uncertainty of $\sim6\,\mathrm{mas}$. The solid cyan lines indicate the fitted outflow trajectories, whereas the dashed cyan lines show these trajectories extrapolated to the maser disk. The orange cross indicates the maser kinematic centre, while the cyan cross indicates the AGN position. The bottom panel shows a schematic representation of the proposed outflow cone geometry. The sketched torus component is based on the B6LB torus morphology in Fig. \ref{fig:mol_morphology}. The grey labels "N" (near) and "F" (far) indicate the respective sides of the warped accretion disk: "N" marks the near side that is in the line of sight of the AGN, while "F" corresponds to the far side that is in shadow, as discussed in Sect. \ref{sec:anisotropic}. }
    \label{fig:outflow_morphology}
\end{figure}

In the top panel of Fig.~\ref{fig:outflow_morphology}, we compare the pc-scale outflow traced by the 259\,GHz continuum with the dusty outflow cone traced by the mid-IR N-band continuum using the MATISSE interferometer \citep{Isbell_2022}. Since the MATISSE measurements lack absolute astrometry, we match the position of the continuum peak in the N-band to the submillimetre continuum peak. We find that the N-band morphology is very similar, suggesting that the dusty outflow cone is cospatial with the edge of the ionised outflow cone. Both tracers show a polar extension with a PA of $\sim 130^{\circ}$ and a more compact horizontal bar at $\mathrm{PA}\sim 92^{\circ}$ surrounding the unresolved continuum peak. We therefore suggest that the clockwise rotation of the contours in our 259\,GHz map has the same origin as the horizontal bar in the N-band \citep{Isbell_2022}: we predominantly detect one edge of each outflow cone (the W edge of the NW cone and the E edge of the SE cone). Since the emissivity of free-free radiation scales as $j_{\mathrm{FF}}\propto n_{\mathrm{e}}^2\times T_{e}^{-0.323}$ \citep{Draine_2011}, and given that $T_{\mathrm{e}} = (0.8$–$3)\times10^4\,\mathrm{K}$, temperature variations alone can account for at most a factor of $\sim1.5$ in flux density contrast. The observed asymmetry therefore likely reflects a difference in electron density between the two edges of the cone, rather than a temperature difference.

Although the morphology of the N band is essentially point-symmetric\footnote{We ascribe the lack of asymmetry in the N-band map from \citet{Isbell_2022} to a bias towards point symmetry in the MATISSE imaging since \citet{Tristram_2014} found polar asymmetry in the N-band.}, the extended emission in the B6LB map at $r>1\,\mathrm{pc}$ to the north-west is only one-sided. Since the outflow appears point-symmetric in B6SB (Fig. \ref{fig:cont_morphology} top panel), we conclude that we detect both ionised outflow cones but that the north-western cone is brighter at the pc scale, making only this cone detectable in the long-baseline map at $r>1\,\mathrm{pc}$. This asymmetry might be the result of slightly denser material in the NW cone. Similarly, \citetalias{Izumi_2023} detected the ionised outflow as one-sided extended emission toward the north-west in H36$\alpha$. 

In the middle panel of Fig.~\ref{fig:outflow_morphology}, we compare the B6LB continuum at the highest spatial resolution ($14.3\times12.2\,\mathrm{mas}^2$ or $0.29\times0.24\,\mathrm{pc}^2$) with the 22\,GHz very long baseline interferometry (VLBI) maser positions of \citet{McCallum_2009} (as described in Sect.~\ref{sec:VLBI_astrometry}). At the highest angular resolution, we resolved the horizontal bar into a curved zigzag structure consisting of two linear outer structures with the same orientation as the edge of the dusty outflow cone and an inner unresolved core. We find that the outflow masers extend farther outwards from the polar axis (and closer to the midplane) than the 259\,GHz and N-band continuum at the northern (southern) side of the eastern (western) outflow edge. This implies a stratification between the ionised outflow and the outflowing maser clumps (blue and purple cones in Fig.~\ref{fig:outflow_morphology}), with the molecular outflow having a slightly wider opening angle.

\subsubsection{Launching or collimation radius}
The curved zigzag morphology of the 259\,GHz continuum indicates that we resolved the base of the ionised outflow cone for the first time in the Circinus AGN. If we assume that the outflowing material follows straight trajectories after leaving the disk, we can estimate the radius of the base as follows. We fit two straight lines to the outer linear structures (solid cyan lines in Fig. \ref{fig:outflow_morphology}) and then extrapolate these trajectories inward to the maser disk (dashed cyan lines in Fig.~\ref{fig:outflow_morphology}). We find that they connect to the disk at the edges of the inner warped maser disk section at $r\approx 0.16\,\mathrm{pc}$ with half-opening angles of $\theta_{\mathrm{W}}\approx44^{\circ}$ and $\theta_{\mathrm{E}}\approx29^{\circ}$ for the western and eastern trajectories respectively\footnote{As measured from the polar axis, which we assume to be perpendicular to the inner section of the maser disk.}. We can interpret this radius in two ways: either as a launching radius or as a collimation radius.

The collimation of AGN outflows has previously been suggested to be caused by the torus or by a warped accretion disk \citep{Fischer_2013}. Since the outer sections of the maser disk (red and blue lines in the bottom panel of Fig.~\ref{fig:outflow_morphology}) coincide with the edges of the ionisation cones, we suggest that the warped maser disk collimates the ionised outflow before the outflow reaches the torus. This cannot be concluded for the "outflow masers", which trace a cone that is wider than the warped maser disk. This is consistent with the molecular outflow being instead collimated by the torus or these maser clumps tracing molecular torus material that is shocked by the outflow. If the observed geometry is indeed the result of collimation, any mechanism with a launching radius $r<0.16\,\mathrm{pc}$ is consistent with our observations.

An alternative interpretation is that we directly observe the launching radius. In the radiation-driven mechanism proposed by \citet{Honig_2017}, the ionised and dusty wind is launched near the dust sublimation radius, which is equal to $r_{\mathrm{sub}}=0.05-0.2\,\mathrm{pc}$ for Circinus (depending on the dust composition and AGN luminosity; \citealt{Isbell_2023}). Our observations are therefore consistent with the outflow mechanism proposed by \citet{Honig_2017}. The assumption of $r_{\mathrm{launch}}=0.16\,\mathrm{pc}$ in \citet{Andonie_2022} is also consistent with our result, while $r_{\mathrm{launch}}=0.53\,\mathrm{pc}$ in \citet{Vollmer_2018} appears too large.

\subsubsection{Anisotropic launching mechanism}\label{sec:anisotropic}
In the mid-IR, the asymmetry observed between the edges of the dusty outflow cones has previously been attributed to anisotropic illumination by a tilted inner accretion disk \citep{Tristram_2014}, with the W and E edges appearing hotter than the N and S edges due to preferential exposure to the central radiation source (see also \citealt{Stalevski_2017, Stalevski_2019}). However, our observations suggest a difference in density, rather than temperature, between both edges of the ionisation cone.

This leads us to propose the following explanation. The continuum morphology at 259\,GHz shows that the outflow trajectories (cyan lines in Fig.~\ref{fig:outflow_morphology}) intersect regions located $\sim0.1\,\mathrm{pc}$ above the outer parts of the warped maser disk. The material located on these near sides of the maser disk (labelled 'N' in the bottom panel of Fig. \ref{fig:outflow_morphology}) lies in direct sight of the AGN. This material can therefore be efficiently launched either by radiation pressure from the inner accretion disk or entrainment in a wide-angle wind that originates close to the AGN. In contrast, material on the far sides of the maser disk (labelled 'F') is shielded by the warped optically thick accretion disk. We therefore suggest that we detect an anisotropic wind launching from a warped disk rather than an anisotropic illumination of an otherwise isotropic wind. Unlike the model of \citet{Stalevski_2017}, our interpretation does not require a tilt of $40^{\circ}$ of the inner accretion disk relative to the polar axis within the inner region.

\subsection{Nuclear dust emission}\label{sec:Phoenix}
\begin{figure}[ht]
    \centering
    \includegraphics[width=\linewidth]{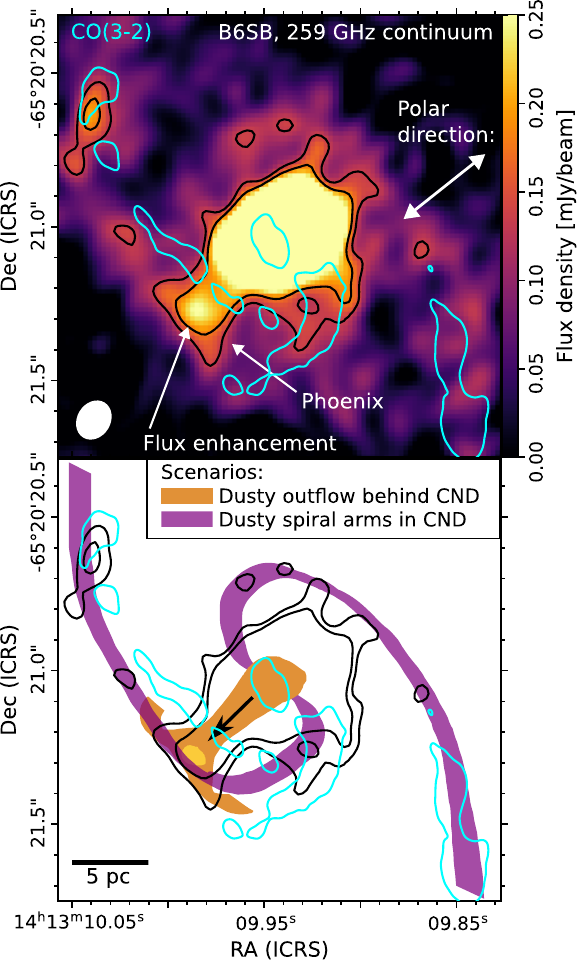}
    \caption{Top panel: The Phoenix feature as observed in the B6SB 259 GHz continuum compared to the B7 CO(3-2) molecular tracer (blue contour at $1.0\,\mathrm{mJy/beam\,km/s}$). We observe a mismatch between the continuum and CO(3-2) morphology, which either implies that the Phoenix feature is associated with a polar outflow perpendicular to the CND or that the inner spiral arm structure traced by dust differs from the molecular spirals. Bottom panel: Illustration of both of these scenarios. The spiral pattern drawn in purple is based on the structure observed in the right panel of Fig. \ref{fig:FF_dust_model}. The B6SB 259 GHz continuum contours are drawn at $[6, 8]\,\times\,\mathrm{rms}$, with $\mathrm{rms}=0.022\,\mathrm{mJy/beam}$.}
    \label{fig:phoenix}
\end{figure}

To the south-east of the nucleus, we detect the Phoenix feature. Its emission, which extends out in the radial direction to a projected distance of $\sim8\,\mathrm{pc}$ from the nucleus, is associated with dust as shown in Sect.~\ref{sec:continuum_origin}. The phoenix ends in a flux enhancement, where it splits into two "wings". When comparing the morphology of the Phoenix with that of the molecular spiral arms, as shown in Fig.~\ref{fig:phoenix}, we find that they trace very different structures. The molecular maps show no structure extending radially outwards from the AGN, and the eastern ridge in the CND does not coincide with the ridge created by the two wings of the Phoenix. We discuss two possible scenarios in the following subsections.

\subsubsection{Outflow scenario}
Because of the morphological discrepancy and the polar orientation, we suggest that the Phoenix might not be part of the CND but is instead located behind the disk as part of a dusty outflow (as illustrated in Fig.~\ref{fig:phoenix}). This dust would not be shielded by the torus, so we expect it to be warmer than the dust in the midplane.

At the kpc scale, \citet{Elmouttie_1998} discovered a bipolar outflow in the $20\,\mathrm{cm}$ radio continuum. The PA of the Phoenix feature ($\approx135^{\circ}$) differs slightly from that of the SE kpc-scale outflow ($\approx115^{\circ}$), but it is nearly perpendicular to the PA of the inner tilted maser disk ($\approx48^{\circ}$) from which it would be launched. It is also almost exactly opposite the direction of the collimated ionised outflow on the NW side ($\approx320^{\circ}$; \citealt{Kakkad_2023}). Therefore, the Phoenix would be the SE counterpart of the NW outflow. The $\sim40\,\mathrm{pc}$-scale warm dust component observed in the mid-IR by \citet{Stalevski_2017} has a significantly different PA of $\approx100^{\circ}$, meaning that it would have to trace a different (temperature) component than the Phoenix.

The fact that the Phoenix abruptly ends in a flux enhancement and splits into two perpendicular wings could imply that the material slows down and eventually falls back to the disk. This would mean that we would directly detect the dust fountain as proposed by \citet{Wada_2016}. Such a backflow of dust and gas would add to the turbulent motion in the disk, which is one proposed mechanism to keep the torus geometrically thick. The flux enhancement could then be associated with a collision of the outstreaming column of material onto a dense clump, causing a build-up of material or heating of the dust. A very similar picture of a collimated outflow splitting into two branches due to a dense clump was presented by \citet{Kakkad_2023} on the NW side of the Circinus AGN (referred to as the "tuning fork") to explain the morphology in the [O\,{\sc iii}] cone.

\subsubsection{Spiral arm scenario}
Another explanation for the mismatch in morphology is that the dusty and molecular arms are simply not cospatial. This offset between dust and molecular gas has been observed at the kpc scale in several spiral galaxies, where this offset results from warm dust that is heated by star formation triggered in the density wave \citep{Patrikeev_2006, Chandar_2017, Vallee_2020}. Because a similar offset is also observed to the NW of the CND between the spiral dust arm and the NW molecular hook, a similar offset at the SE side of the CND would be plausible.

\subsection{CND kinematics}\label{sec:CND_kinematics}
\begin{figure*}[h]
    \centering
    \includegraphics[width=\linewidth]{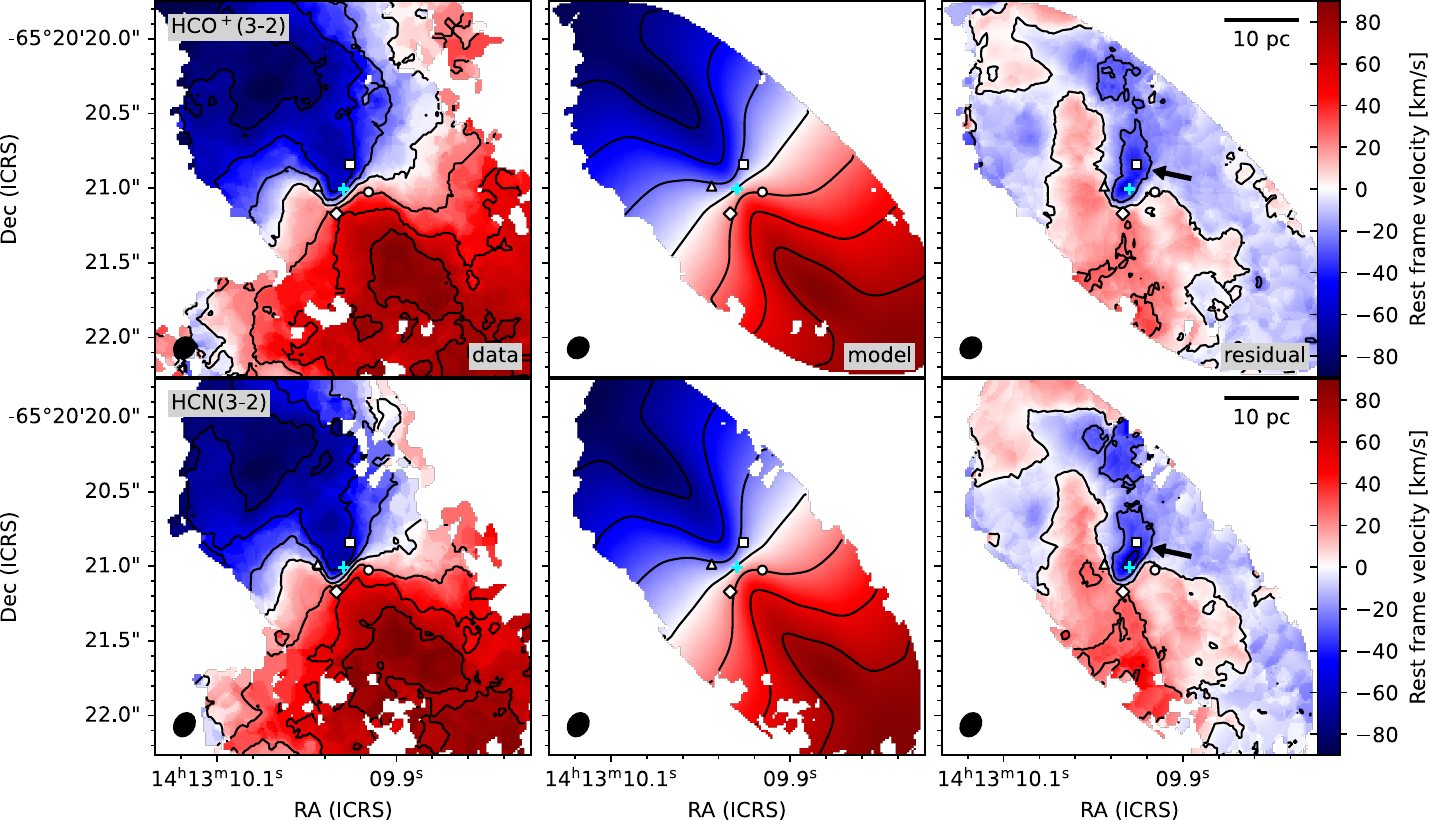}
    \caption{Velocity fields of the CND for HCO$^+$(3-2) (top) and HCN(3-2) (bottom) from the B6SB dataset as observed (left), as modelled using $^\mathrm{3D}$BAROLO (middle) and the residual velocity field that results from subtracting the model from the data (right). The black arrows in the right panels indicate the position of the blue residual feature that coincides with the position of the NW hook. Velocity contour levels are drawn at $[-75, -50, -25, 0, 25, 50, 75]\,\,\mathrm{km\,s^{-1}}$. The rest frame velocity is defined relative to $V_{\mathrm{sys}}=441.4\,\mathrm{km\,s^{-1}}$. Indicated on all maps are the point apertures used to extract the 1-D kinematic profiles shown in Fig. \ref{fig:1D_kin_profiles} using corresponding symbols in white. The synthesised beam sizes are drawn in the bottom left as filled ellipses, and the cyan cross indicates the AGN position.}
    \label{fig:BBAROLO_velfield}
\end{figure*}

The material surrounding molecular tori is hypothesised to contain non-circular motions responsible for inflows and outflows, for example, as part of the fountain model introduced by \citet{Wada_2016}.  We are therefore interested in the kinematics of the molecular gas in the Circinus CND. In particular, the spiral structure observed in the CND could induce non-circular motions in the material orbiting the SMBH \citep{Roberts_1979, VandeVen_2010}. Moment-one maps of the B6SB HCO$^+$(3-2) and HCN(3-2) emission are presented in the left panels of Fig.~\ref{fig:BBAROLO_velfield}. We find a strong dipole that indicates that the kinematics is dominated by rotation. However, on the basis of the velocity contours, we find that the observed velocity field is not symmetric around the minor and major rotation axes of this dipole. The CND must therefore contain gas motions that deviate significantly from a pure circular rotation around a single rotation axis.

To determine whether the observed asymmetry arises from non-circular gas movements or a warp in the disk position angle (PA), we modelled the HCO$^+$(3–2) and HCN(3–2) kinematics using the software $^\mathrm{3D}$BAROLO \citep{BBAROLO_2015}, which fits 3D tilted ring models directly to emission line data cubes. Initially, we only allowed circular motion in the $^\mathrm{3D}$BAROLO model such that any non-circular motion can be investigated from the residual velocity field after subtracting the model from the observed velocity field. We defined 29 tilted concentric rings placed at regular intervals between $r=0.05^"$ and $r=1.45^"$ centred on the AGN position. For each ring, we fitted the rotation velocity $V_{\mathrm{rot}}$, the dispersion velocity within the disk $V_{\mathrm{disp}}$, the inclination angle $i$, the position angle of the major axis on the receding side $\mathrm{PA}$ and the systemic velocity $V_{\mathrm{sys}}$ using $\chi^2$ minimisation. We used a pixel-by-pixel normalisation of the surface density, which allowed the model to reproduce the clearly non-symmetric intensity distribution in the Circinus CND. The fitting process was executed in two stages. In the first stage, all free parameters were unconstrained and were fitted to each ring separately. In the second stage, the inclination and position angles were regularised through a radial Bezier function, and the systemic velocity was fixed at the median value of all rings. In addition, we performed an alternative fit in which the radial velocity $V_{\mathrm{rad}}$ is included as a free parameter in the model. We discuss this additional fit in Sect. \ref{sec:Vrad_free}.

\begin{figure}[h]
    \centering
    \includegraphics[width=\linewidth]{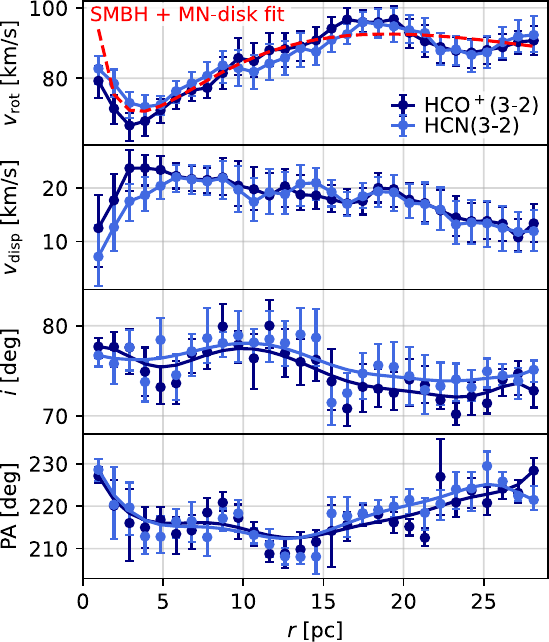}
    \caption{Fitted kinematic and morphological radial parameters of the CND from $^\mathrm{3D}$BAROLO modelling of the HCO$^+$(3-2) and HCN(3-2) transitions in the B6SB dataset. The top two panels show the $^\mathrm{3D}$BAROLO fitting results and the associated uncertainties after the second fitting iteration, while the bottom two panels show the fit results and uncertainties from the first iteration as scatter points with the regularised radial parameter profiles overlaid as solid curves that are used in the second iteration. The SMBH + Miyamoto-Nagai disk rotation curve fit is overlaid in the top panel as a red dashed line.}
    \label{fig:BBAROLO_params}
\end{figure}

The resulting radial parameter profiles are shown in Fig.~\ref{fig:BBAROLO_params}, and the resulting moment-one model and residual velocity fields are shown in Fig.~\ref{fig:BBAROLO_velfield}. We discuss the major-axis position-velocity (PV) diagram of both the model and the data in Sect. \ref{sec:PV_diagram}. In general, we find that the fitted parameters and the residual velocity fields are robust against changes in initial parameters. We also find that HCO$^+$(3-2) and HCN(3-2), within the statistical uncertainties, trace the same circular rotation structure in the disk. The systemic velocities are fitted to be $441.9\pm0.9\,\mathrm{km\,s^{-1}}$ and $440.8\pm0.9\,\mathrm{km\,s^{-1}}$ for the HCO$^+$(3-2) and HCN(3-2) transitions, respectively. We therefore conclude that the average fitted systemic velocity of the CND is equal to $V_{\mathrm{sys}}=441.4\pm0.7\,\mathrm{km\,s^{-1}}$, which is consistent with the nominal systemic velocity of $439\pm2\,\mathrm{km\,s^{-1}}$ of the galaxy and the nuclear molecular material \citep{Freeman_1977, Curran_1998}. We find $V_{\mathrm{disp}}$ to range between $\approx10\,\mathrm{km\,s^{-1}}$ and $\approx23\,\mathrm{km\,s^{-1}}$ throughout the disk, and the inclination angle and position angle are $i=75\pm3^{\circ}$ and $PA=218\pm6^{\circ}$. These results are consistent with the kinematic analysis from \citetalias{Tristram_2022}, but have smaller uncertainties.

The rise in the rotation curve for $r<3\,\mathrm{pc}$ (see Fig. \ref{fig:BBAROLO_params}) indicates a mass concentration near the disk centre. Although the interpretation of these results warrants caution for the inner two rings (since the minor axis is unresolved, as discussed in Sect.~\ref{sec:PV_diagram}), the observed increase in $V_{\mathrm{rot}}$ and the drop in $V_{\mathrm{disp}}$ at $r<3\,\mathrm{pc}$ are consistent with the higher resolution modelling by \citetalias{Izumi_2023}. Except for the innermost ring, all fitted line-of-sight velocities are higher than would be expected from the Keplerian profile of a $M_{\bullet}=\left(1.7\pm0.3\right)\times10^6\,\mathrm{M}_{\odot}$ SMBH (see Fig. \ref{fig:BBAROLO_PV}; \citealt{Greenhill_2003}). Approximating with spherical symmetry, the dynamical mass enclosed within radius $r$ is given by: $M(r)= r V_{\mathrm{rot}}(r)^2/\mathrm{G}$, with $\mathrm{G}$ the gravitational constant. Using the fitted rotational velocity of $V_{\mathrm{rot}}\approx70\,\mathrm{km\,s^{-1}}$ at $r=2\,\mathrm{pc}$ (which encloses the torus structure), we find a dynamical mass of $\approx2.6\times10^6\,\mathrm{M}_{\odot}$. This implies a dynamical torus mass of $\sim9\times10^5\,\mathrm{M}_{\odot}$ after subtraction of the SMBH mass\footnote{We ignore the mass of the accretion disk, on which \citet{Greenhill_2003} placed an upper limit of $M_{\mathrm{acc}}\lesssim4\times10^5\,\mathrm{M}_{\odot}$ within $r=0.4\,\mathrm{pc}$.}. At larger radii, the rotational velocity increases to $\sim90\,\mathrm{km\,s^{-1}}$ out to a radius of 17 pc, implying that the CND kinematics is dominated by self-gravitation. We model this rotation curve using a $1.7\times10^6\,\mathrm{M}_{\odot}$ point mass surrounded by a Miyamoto-Nagai (MN) disk \citep{Miyamoto_1975}, for which we assume a scale height\footnote{Since $V_{\mathrm{disp}}/V_{\mathrm{rot}}\approx0.22$. We follow the method from \citealt{Krolik_1988}, which states that $b/a\sim V_{\mathrm{disp}}/V_{\mathrm{rot}}$ in the torus disk. If we, for example, instead choose $b/a=0.1$, $M_{\mathrm{CND}}$ is reduced by 7\%.} of $b/a=0.22$ (with $a$ and $b$ the length and thickness scales of the MN disk) and fit a disk scale length of $a=13.3\pm0.5\,\mathrm{pc}$ and an enclosed disk mass of $M_{\mathrm{MN}}(r\leq28\,\mathrm{pc})\equiv M_{\mathrm{CND}}=\left(3.2\pm0.2\right)\times10^7\,M_{\odot}$ (see Fig. \ref{fig:BBAROLO_params}).

\begin{figure}[h]
    \centering
    \includegraphics[width=\linewidth]{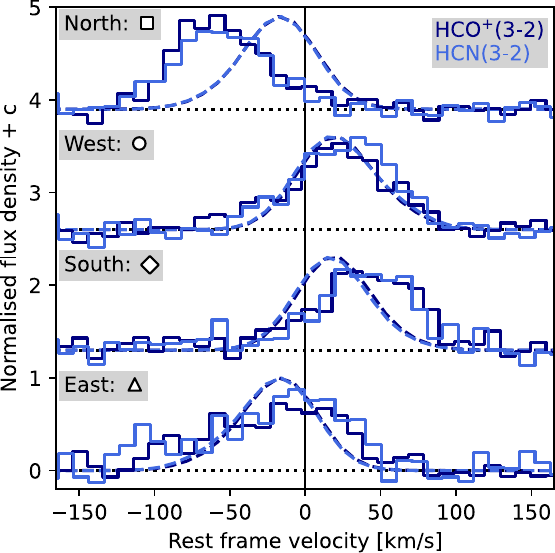}
    \caption{Kinematic 1-D profiles of HCO$^+$(3-2) and HCN(3-2) in B6SB extracted from the four point apertures as indicated in Fig. \ref{fig:BBAROLO_velfield} using corresponding markers. We show the data as a step curve and the $^\mathrm{3D}$BAROLO model as dashed curves. The 1-D kinematic profiles are normalised by the maximal flux density in the model. The spectra are offset by a term $c$ as indicated by the horizontal dotted lines.}
    \label{fig:1D_kin_profiles}
\end{figure}

Looking at the residuals in the velocity fields (right panels in Fig. \ref{fig:BBAROLO_velfield}), we find that large residuals remain after subtraction of the model. This is a clear indication of non-circular motions. The blue residual to the north-west of the SMBH (indicated with a black arrow in Fig. \ref{fig:BBAROLO_velfield}) is the strongest and has a magnitude of $\sim-40\,\mathrm{km\,s^{-1}}$. In addition, we observe a significant redshift residual in a more extended region to the south-east of the SMBH with a magnitude of $\sim20\,\mathrm{km\,s^{-1}}$. As discussed in Sect. \ref{sec:improper_modelling}, these residuals are not caused by improper modelling (e.g., due to the PA being significantly larger in the inner disk than assumed by the model). In Fig. \ref{fig:1D_kin_profiles}, we show four 1-D kinematic profiles that we extracted from the point apertures indicated in Fig. \ref{fig:BBAROLO_velfield}. We find that the blue and red residuals in the northern and southern apertures are not caused by a secondary blue or red shifted velocity component. Instead, we find that all apertures show an approximately Gaussian profile that is similar to the model but shifted to a different velocity. Therefore, these residuals are the result of motions within the disk itself and not due to an additional molecular outflow that is launched from the disk or material falling back onto the disk.

\subsection{Torus feeding through inflow in the CND}\label{sec:inflow}
The blue-shifted emission in the residual maps of Fig.~\ref{fig:BBAROLO_velfield}, located north-west of the SMBH, originates, at least for the most part, from the structure we identified as the NW hook. This is because in the HCO$^+$(3-2) and HCN(3-2) maps of B6SB, on which the kinematic modelling is based, the blue residual spatially coincides with the NW-arm (cyan contours in Fig.~\ref{fig:mol_morphology}). Further confirmation comes from overlaying the blue residual onto the higher-resolution band 7 maps in Figure~\ref{fig:mol_morphology_B7}, where the blue residual again coincides with the NW hook.

Since the NW hook is located on the far side of the CND, we interpret this blueshift as an inflow along the NW hook. We interpret this as an inflow rather than an outflow because (a) the corresponding 1-D kinematic profile only shows one blue-shifted velocity component from the disk rather than a disk component and an extra blue-shifted outflow component (b) the blue-shifted emission originates from a structure that morphologically resembles a spiral arm, which would be associated with instabilities within the self-gravitating gaseous disk and (c) the position angle and location of the known outflow, which is tilted towards the west side of the rotation axis (see Figs. \ref{fig:cont_morphology}, \ref{fig:FF_dust_model}, \ref{fig:outflow_morphology} and the dusty and ionised outflow as traced by \citealt{Stalevski_2017, Isbell_2022, Kakkad_2023} and \citetalias{Izumi_2023}), do not match with the position of the blue residual, which is located to the north-east of the rotation axis. Therefore, we conclude that we observe gas that is being transported inward along the NW hook, which connects to the approaching side of the torus, thereby feeding the torus.

Following the same logic as for the NW-hook, the red-shifted feature on the near side of the CND would also be associated with an inflow. Based on the morphology of the residual velocity map, this inflow would be less collimated and smaller in magnitude ($\sim20\,\mathrm{km\,s^{-1}}$). Since this red-shifted region coincides with the location of the E ridge and the SW hook, these southern spiral structures could be responsible for the angular momentum transfer required for the inward transport of material on this side of the disk. However, in contrast to the inflow motion observed to the NW, this inward motion would be perpendicular to the E-ridge rather than along the length of the spiral arm. At the location where the SW hook connects to the torus, we observe no significant velocity residual. This does not imply that there is no inward transport of material on this side of the torus because we are insensitive to a radial inflow at this location (which would be perpendicular to our line of sight).

As an extra check that the observed residuals imply inflow, we performed a second $^\mathrm{3D}$BAROLO fit where we allow the rings to have non-zero axisymmetric radial velocities (see Sect. \ref{sec:Vrad_free}). We find that the model consistently predicts an inflow in the central region of the CND ($r<15\,\mathrm{pc}$) that increases in velocity towards the SMBH, with a mean magnitude of $V_{\mathrm{rad}}\sim-12\,\mathrm{km\,s^{-1}}$ between 5\,pc and 15\,pc (see Fig.~\ref{fig:BBAROLO_params_Vradfree}). The resulting residuals on the velocity map are similar in shape but have a smaller magnitude than those of the model that fixes $V_{\mathrm{rad}}=0\,\mathrm{km\,s^{-1}}$ (Fig. \ref{fig:BBAROLO_velfield_Vradfree}). This is not surprising, since we measured a projected inflow velocity of $\approx40\,\mathrm{km\,s^{-1}}$ in the NW hook, which is much faster than the inflow in the axisymmetric inflow model. We therefore conclude that the inflow is not axisymmetric and is dominated by the motions in or near the spiral arms.

Nuclear spirals are known to exist on $>10\,\mathrm{pc}$ scales in AGN-bearing galaxies. These are often connected to the CND\footnote{In some cases, such as \citet{Combes_2019}, this disk is referred to as the torus, where it is uncertain if this structure is what obscures the AGN. By that definition, the disk we refer to as the CND could be called the torus, implying that we observe spiral structure within the torus.} and thought to be responsible for AGN feeding based on the non-axisymmetric potential that they produce (see e.g., \citealt{Pogge_2002, Prieto_2005, Combes_2014, Audibert_2019, Combes_2019, Storchi_Bergmann_2019}). Our observations of the Circinus nucleus reveal a case where we observe a clear spiral structure within the CND that connects to a kinematically distinct substructure (the torus) and transports material down to the pc-scale torus. Such a spiral structure in the Circinus CND is not unexpected, since the Toomre-Q parameter is smaller than unity outside $r=1\,\mathrm{pc}$ (\citealt{Toomre_1964}; \citetalias{Izumi_2023}), meaning that we expect gravitational instabilities to grow into spiral structures. These bars and spirals then produce negative torques that drive inflow \citep{Buta_1996, Combes_2021}.

\subsubsection{Torus feeding rate}\label{sec:inflow_rate}
The feeding rate of the torus through the NW hook can be estimated by multiplying the radial inflow velocity by the linear gas mass density in the inner straight bar of the hook. Since the stars in the CND are collisionless, their motion is expected to be decoupled from that of the gas \citep{Muller_2011, Schartmann_2017}. We thus assume that only the gas contributes to the inflow rate. Because the disk is tilted by $i=76\pm1^{\circ}$ at the point aperture in the NW hook (square marker in Fig.~\ref{fig:BBAROLO_velfield}), it intersects the CND at a deprojected distance of $r=14\,\mathrm{pc}$. At this aperture, we measure a residual velocity of $V_{\mathrm{LOS, \,NW}}=-39\pm2\,\mathrm{km\,s^{-1}}$. Assuming that the residual gas motion is purely radial after subtraction of the fitted rotational velocity component, we find a de-projected radial velocity of $V_{\mathrm{rad,\,NW}}=V_{\mathrm{LOS}}/\left(\sin i\times\cos \phi\right)=-43\pm2\,\mathrm{km\,s^{-1}}$. Here, $\phi=23^{\circ}$ is the angle between the inner bar of the NW hook and the disk minor axis.

It is difficult to determine the gas mass enclosed in the NW hook from the line intensities. This is because, assuming a gas fraction of $f_{\mathrm{gas}}\sim0.5$ \citep{Mueller_2006, Hicks_2009}, the Miyamoto-Nagai disk model predicts a mean face-on column density of $N_{\mathrm{H}_2}=4.8\times10^{23}\,\mathrm{cm}^{-2}$ at $r=14\,\mathrm{pc}$, which makes all observed transitions optically thick\footnote{Using \texttt{RADEX} \citep{Radex_2007}, we find $\tau\gg1$ for $T_{\mathrm{gas}}\leq50\,\mathrm{K}$, line widths of $50\,\mathrm{km\,s^{-1}}$, $n_{H_2} = n_{\mathrm{crit}}$ for each transition and chemical abundances of $X_{\mathrm{CO}}\sim 10^{-4}$ and $X_{\mathrm{HCN}}\sim X_{\mathrm{HCO^+}}\sim 3\times10^{-9}$ \citep{Viti_2014, Laszlo_2016, Rosas_2025}.}. We therefore have to rely on more rudimentary estimates. Since the width of the SW hook is partially resolved in B7 (see Fig. \ref{fig:mol_morphology_B7}), we assume a diameter of $D\sim1\,\mathrm{pc}$ for the NW hook, comparable to the B7 beam size. Approximating its shape as a cuboid of side length $D=1\,\mathrm{pc}$, elongated radially along the inner straight bar, and filled with gas at a mean density of $n_{\mathrm{H}_2}\sim n_{\mathrm{crit,\,HCO^+(3-2)}}=10^6\,\mathrm{cm^{-3}}$, we find an inflow rate of:

\begin{equation}\label{eq:inflowNW_lindens}
    \begin{aligned}
    \dot{M}_{\mathrm{NW-hook}}=-V_{\mathrm{rad,\,NW}} \times m_{\mathrm{H}_2}\times n_{\mathrm{H}_2}\times D^2\sim2.2\,\mathrm{M}_{\odot}\mathrm{yr}^{-1},
    \end{aligned}
\end{equation}

where $m_{\mathrm{H}_2}=3.35\times10^{-27}\,\mathrm{kg}$ is the mass of a hydrogen molecule. In this toy model, the NW hook has a face-on column density of $N_{\mathrm{H}_2}=n_{\mathrm{H}_2}\times D=3.1\times10^{24}\,\mathrm{cm}^{-2}$. If we instead set $N_{\mathrm{H}_2}=4.8\times10^{23}\,\mathrm{cm}^{-2}$ equal to the mean column density of the Miyamoto-Nagai disk at $r=14\,\mathrm{pc}$ (implying $n_{\mathrm{H}_2}=1.5\times10^{5}\,\mathrm{cm}^{-3}$), we find an inflow rate of $\dot{M}_{\mathrm{NW-hook}}\sim0.3\,\mathrm{M}_{\odot}\mathrm{yr}^{-1}$. We can treat this latter value as a lower limit because the gas density in the NW hook is likely higher than the average density throughout the CND.

To determine the total feeding rate of the torus, we also have to consider the mass inflow through the rest of the disk. Since an inflow along the SW hook would be perpendicular to our line of sight, we cannot perform an analogous calculation of $\dot{M}_{\mathrm{SW-hook}}$. However, based on the second $^\mathrm{3D}$BAROLO fit, where $V_{\mathrm{rad}}$ was left as a free parameter to describe the axisymmetric radial velocities in the rings, we can calculate an inflow rate for the entire disk (for details, see App. \ref{sec:mass_inflow_profile}). The resulting radial inflow profile (bottom panel Fig.~\ref{fig:BBAROLO_params_Vradfree}) is constant at $\dot{M}_{\mathrm{inflow}}=7.5\pm1.5\,\mathrm{M}_{\odot}\mathrm{yr}^{-1}$ between $r=3\,\mathrm{pc}$ and $r=14\,\mathrm{pc}$. As discussed in App. \ref{sec:mass_inflow_profile}, we expect this total inflow rate to be an overestimate. We therefore constrain the torus feeding rate to $\dot{M}_{\mathrm{feed}}=0.3-7.5\,\mathrm{M}_{\odot}\mathrm{yr}^{-1}$.

These feeding rates are higher than those reported by \citetalias{Izumi_2023} for the Circinus AGN, who derived $\dot{M}_{\mathrm{feed}}=0.20-0.35\,\mathrm{M}_{\odot}\mathrm{yr}^{-1}$ from the redshifted absorption profile described in Sect. \ref{sec:results_molecular}. Our estimates are likely more robust, as \citet{Baba_2024} showed that infalling clumps traced by absorption do not necessarily provide a reliable measure for the total mass inflow rate. Furthermore, \citetalias{Izumi_2023} used $r_{\mathrm{inflow}}=0.27\,\mathrm{pc}$ in their calculation, whereas \citet{Baba_2024} showed that such absorption signatures are more likely to originate at much larger radii in the CND. Our inferred feeding rate also exceeds that obtained in the hydrodynamic simulation by \citet{Baba_2024}, which yielded a maximum of $\dot{M}_{\mathrm{feed}}\sim0.35$ at $r=5\,\mathrm{pc}$.

\subsubsection{Mass flow and feeding timescales}
The Eddington accretion rate of the Circinus AGN is equal to $\dot{M}_{\mathrm{acc}}=L_{\mathrm{Edd}}/c^2\eta\approx0.04\,\mathrm{M}_{\odot}\mathrm{yr}^{-1}$ (for a radiative efficiency of $\eta=0.1$ and an Eddington luminosity of $L_{\mathrm{Edd}}\approx5.6\times10^{10}\,\mathrm{L}_{\odot}$, \citealt{Tristram_2007}), which is only 12\% of our lower torus feeding limit. This implies that the majority (>88\%) of the incoming mass must be removed from the AGN before reaching the accretion disk and must be expelled to distances beyond $r\sim14\,\mathrm{pc}$ as part of an outflow. Estimates for the ionised and molecular outflows range between $\dot{M}_{\mathrm{out,\,ion}}=0.01-0.22\,\mathrm{M}_{\odot}\mathrm{yr}^{-1}$ (\citealt{Fonseca_2021, Kakkad_2023}; \citetalias{Izumi_2023}) and $\dot{M}_{\mathrm{out,\,mol}}=0.35-12.3\,\mathrm{M}_{\odot}\mathrm{yr}^{-1}$ \citep{Zschaechner_2016}, respectively. Our values for the torus feeding rate are thus consistent with those required to sustain accretion and outflow. Although part of the inflow is used for star formation, this is a small fraction of the total mass flow since \citet{Esquej_2014} reported a rate of only $\dot{M}_{\mathrm{SFR}}=0.13\,\mathrm{M}_{\odot}\mathrm{yr}^{-1}$ within $75\,\mathrm{pc}$ from the AGN.

Since we estimated dynamical masses of $M_{\mathrm{torus}}\sim9\times10^5\,\mathrm{M}_{\odot}$ and $M_{\mathrm{CND}}=\left(3.2\pm0.2\right)\times10^7\,M_{\odot}$, the implied timescales are $t_{\mathrm{feed,\,torus}}=120\,\mathrm{kyr}-2.7\,\mathrm{Myr}$ and $t_{\mathrm{depl,\,CND}}=4.3-97\,\mathrm{Myr}$. Assuming a steady state with approximately $\dot{M}_{\mathrm{feed}}\approx\dot{M}_{\mathrm{acc}}+\dot{M}_{\mathrm{outflow}}=\dot{M}_{\mathrm{depl}}$, we also expect $t_{\mathrm{depl,\,torus}}\sim t_{\mathrm{feed,\,torus}}$. Since both the torus and the CND act as mass reservoirs for the AGN, any variation in the activity of the AGN caused by inhomogeneities in the CND would be expected on timescales of $\sim t_{\mathrm{depl,\,CND}}$ and would be time-smoothed on timescales larger than $t_{\mathrm{depl,\,torus}}$. These timescales are similar to those observed for the (de)activation of AGN, both based on fluctuations in multi-phase outflows (e.g., $\sim3\,\mathrm{Myr}$ in the Fornax A galaxy, \citealt{Maccagni_2021}) and on radio jets, which typically show AGN activity cycles lasting $1-100\,\mathrm{Myr}$ \citep{Parma_1999, Brienza_2017, Combes_2021}. This suggests that such fluctuations in AGN may be directly related to the variability in torus feeding by the CND.

\section{Conclusions}\label{sec:conclusion}
In this work, we use high-resolution (down to $13\,\mathrm{mas}$ or $0.25\,\mathrm{pc}$) ALMA observations of the Circinus AGN to investigate the feeding of the molecular torus and the launching of the ionised outflow. We focus on the CND (size $\sim50\,\mathrm{pc}$), the torus (size $\sim2.4\,\mathrm{pc}$), and the AGN outflows at pc scale. We investigated the emission mechanisms and morphology of the continuum radiation between $86\,\mathrm{GHz}$ and $665\,\mathrm{GHz}$ and studied the morphology and kinematics of the molecular gas using the HCO$^+$(3-2), HCO$^+$(4-3), HCN(3-2), HCN(4-3) and CO(3-2) emission lines. We summarise our findings as follows.

\begin{itemize}
    \item At $1\,\mathrm{pc}\lesssim r\lesssim28\,\mathrm{pc}$ distance from the nucleus, the millimetre continuum traces thermal emission from cold dust located in the equatorial midplane of the CND and the torus. The cold dust in the CND is distributed in two spiral arms. Toward the SE from the nucleus, we detect the Phoenix feature, which we interpret either as warm dust in an outflowing dust fountain or as cold dust in a spiral structure that is offset from the molecular gas spirals in the CND.
    
    \item At the parsec scale, the nuclear millimetre continuum traces optically thin free-free radiation from the ionised outflow. We derive electron densities on the order of $n_e\sim10^4\,\mathrm{cm}^{-3}$ for $T_{\mathrm{e}}\sim10^4\,\mathrm{K}$. The ionised outflow cone morphology matches that of the dusty outflow cone but differs from the morphology of the molecular outflow, suggesting a stratification between the molecular outflow (wider opening angle) and the dusty and ionised outflow (smaller opening angle).
    
    \item The millimetre continuum shows an unresolved ($\lesssim0.25\,\mathrm{pc}$) central emission peak. We interpret this as optically thin free-free emission from X-ray heated gas in the corona or above the maser disk with $n_{\mathrm{e}}\sim1.3\times10^6\,\mathrm{cm}^{-3}$ for $T_{\mathrm{e}}\sim6\times10^6\,\mathrm{K}$. We do not detect synchrotron emission from a jet or the corona.
    
    \item We measure the base of the ionised outflow to have a radius of $r\approx 0.16\,\mathrm{pc}$ -- which is either a collimation or a launching radius -- and determine half-opening launch angles of $\theta_{\mathrm{W}}\approx44^{\circ}$ and $\theta_{\mathrm{E}}\approx29^{\circ}$. We find evidence that the ionised outflow is collimated by the warped accretion disk and that the ionisation cones appear asymmetric due to an anisotropic launching mechanism (rather than anisotropic illumination).
    
    \item The molecular line emission reveals spiral arms in the CND that consist of four distinct substructures and differ in position from the dusty spirals. Two of these structures, the NW hook and the SW hook, connect to the torus.
    
    \item We find that the kinematics of the CND is dominated by rotation due to self-gravitation with dynamical masses of $M_{\mathrm{CND}}=\left(3.2\pm0.2\right)\times10^7\,M_{\odot}$ and $M_{\mathrm{torus}}\sim9\times10^5\,\mathrm{M}_{\odot}$. In addition, we observe non-circular inward motion on the order of $20-40\,\mathrm{km\,s^{-1}}$ that is concentrated in the spiral structure.
    
    \item These inward gas flows suggest that the non-axisymmetric potential generated by the spiral arms is responsible for driving the inflow from the CND down to the pc-scale torus.
    
    \item We determine the torus feeding rate to be $\dot{M}_{\mathrm{feed}}=0.3-7.5\,\mathrm{M}_{\odot}\mathrm{yr}^{-1}$, which implies that $>88\%$ of the material that flows towards the torus is expelled in outflows before it reaches the inner accretion disk. These rates imply characteristic feeding timescales of $t_{\mathrm{feed,\,torus}}=120\,\mathrm{kyr}-2.7\,\mathrm{Myr}$ and $t_{\mathrm{depl,\,CND}}=4.3-97\,\mathrm{Myr}$, which suggests that the fluctuations in AGN activity often observed on such timescales may be caused by variability in the torus feeding rate.
\end{itemize}

In summary, we report the first resolved detection of the ionised outflow base in Circinus, unveiling important details about the launching mechanism, and we provide the most detailed kinematic evidence to date for torus feeding at parsec scales through spiral arms that are embedded in the CND.
Observations at similar spatial resolution of other AGN will be required to confirm if these mechanisms are universal or specific to the Circinus AGN, and higher-resolution observations in the inner parsec of the Circinus nucleus are needed to constrain the launching mechanism.

\section*{Data availability}
The continuum and molecular moment zero maps presented in Figs. \ref{fig:cont_morphology}, \ref{fig:CO_annotated}, \ref{fig:mol_morphology} and \ref{fig:mol_morphology_B7} as well as an extended version of Table \ref{table:SED_data} are available at the Strasbourg astronomical Data Center (CDS) via anonymous ftp to cdsarc.u-strasbg.fr (130.79.128.5) or via \href{http://cdsweb.u-strasbg.fr/cgi-bin/qcat?J/A+A/}{http://cdsweb.u-strasbg.fr/cgi-bin/qcat?J/A+A/}.

\begin{acknowledgements}
The authors thank B. Vollmer, C. Ricci, A. Pizzetti, S. Viti, J.W. Isbell, Y. Zhang, J.F. Gallimore, J.J. Butterworth, P.P. van der Werf, F. Fraternali, P.E. Mancera Piña, H.J.A. Röttgering, C.A. Vleugels, N.N. Geesink, R.S. Dullaart, F.J. Ballieux, A. Krämer, W.C. Schrier, B. Goesaert, V. Baeten \& D.J. Remmelts for their advice and many helpful discussions. We also thank the anonymous reviewer for the thorough report, which significantly improved the quality of this paper. W.M. Goesaert acknowledges support from ESO and the Leiden University Fund. This paper makes use of the following ALMA data: 2018.1.00581.S, 2017.1.00575.S and 2019.1.00013.S. ALMA is a partnership of ESO (representing its member states), NSF (USA) and NINS (Japan), together with NRC (Canada), MOST and ASIAA (Taiwan), and KASI (Republic of Korea), in cooperation with the Republic of Chile. The Joint ALMA Observatory is operated by ESO, AUI/NRAO and NAOJ.
\end{acknowledgements}

\bibliographystyle{aa}
\bibliography{bibliography}

\begin{thebibliography}{96}
\expandafter\ifx\csname natexlab\endcsname\relax\def\natexlab#1{#1}\fi

\bibitem[{Altenhoff {et~al.}(1960)Altenhoff, Mezger, Wendker, \& Westerhout}]{Altenhoff_1960}
Altenhoff, W., Mezger, P.~G., Wendker, H., \& Westerhout, G. 1960, Veröffentlichungen der Univer-sitäts-Sternwarte zu Bonn, 59, 48

\bibitem[{{Andonie} {et~al.}(2022){Andonie}, {Ricci}, {Paltani}, {Ar{\'e}valo}, {Treister}, {Bauer}, \& {Stalevski}}]{Andonie_2022}
{Andonie}, C., {Ricci}, C., {Paltani}, S., {et~al.} 2022, \mnras, 511, 5768

\bibitem[{Andreani {et~al.}(2018)Andreani, Carpenter, Trigo, Humphreys, Nagai, Remijan, \& Zwaan}]{Andreani_2018}
Andreani, P., Carpenter, J., Trigo, M.~D., {et~al.} 2018, ALMA Cycle 6 Proposer’s Guide, ALMA Doc. 6.2 v1.0

\bibitem[{{Antonucci}(1993)}]{Antonucci_1993}
{Antonucci}, R. 1993, \araa, 31, 473

\bibitem[{{Audibert, A.} {et~al.}(2019){Audibert, A.}, {Combes, F.}, {García-Burillo, S.}, {Hunt, L.}, {Eckart, A.}, {Aalto, S.}, {Casasola, V.}, {Boone, F.}, {Krips, M.}, {Viti, S.}, {Muller, S.}, {Dasyra, K.}, {van der Werf, P.}, \& {Martín, S.}}]{Audibert_2019}
{Audibert, A.}, {Combes, F.}, {García-Burillo, S.}, {et~al.} 2019, A\&A, 632, A33

\bibitem[{{Baba} {et~al.}(2024){Baba}, {Wada}, {Izumi}, {Kudoh}, \& {Matsumoto}}]{Baba_2024}
{Baba}, S., {Wada}, K., {Izumi}, T., {Kudoh}, Y., \& {Matsumoto}, K. 2024, \apj, 966, 15

\bibitem[{Beckmann {et~al.}(2006)Beckmann, Gehrels, Shrader, \& Soldi}]{Beckmann_2006}
Beckmann, V., Gehrels, N., Shrader, C.~R., \& Soldi, S. 2006, The Astrophysical Journal, 638, 642

\bibitem[{{Blundell} \& {Kuncic}(2007)}]{Blundell_2007}
{Blundell}, K.~M. \& {Kuncic}, Z. 2007, \apjl, 668, L103

\bibitem[{{Brienza} {et~al.}(2017){Brienza}, {Godfrey}, {Morganti}, {Prandoni}, {Harwood}, {Mahony}, {Hardcastle}, {Murgia}, {R{\"o}ttgering}, {Shimwell}, \& {Shulevski}}]{Brienza_2017}
{Brienza}, M., {Godfrey}, L., {Morganti}, R., {et~al.} 2017, \aap, 606, A98

\bibitem[{{Buta} \& {Combes}(1996)}]{Buta_1996}
{Buta}, R. \& {Combes}, F. 1996, \fcp, 17, 95

\bibitem[{Chan \& Krolik(2016)}]{Chan_2016}
Chan, C.-H. \& Krolik, J.~H. 2016, The Astrophysical Journal, 825, 67

\bibitem[{Chandar {et~al.}(2017)Chandar, Chien, Meidt, Querejeta, Dobbs, Schinnerer, Whitmore, Calzetti, Dinino, Kennicutt, \& Regan}]{Chandar_2017}
Chandar, R., Chien, L.-H., Meidt, S., {et~al.} 2017, The Astrophysical Journal, 845, 78

\bibitem[{{Combes}(2021{\natexlab{a}})}]{Combes_2021}
{Combes}, F. 2021{\natexlab{a}}, {Active Galactic Nuclei: Fueling and Feedback}

\bibitem[{{Combes}(2021{\natexlab{b}})}]{Combes_2021b}
{Combes}, F. 2021{\natexlab{b}}, in IAU Symposium, Vol. 359, Galaxy Evolution and Feedback across Different Environments, ed. T.~{Storchi Bergmann}, W.~{Forman}, R.~{Overzier}, \& R.~{Riffel}, 312--317

\bibitem[{{Combes} {et~al.}(2019){Combes}, {Garc{\'\i}a-Burillo}, {Audibert}, {Hunt}, {Eckart}, {Aalto}, {Casasola}, {Boone}, {Krips}, {Viti}, {Sakamoto}, {Muller}, {Dasyra}, {van der Werf}, \& {Martin}}]{Combes_2019}
{Combes}, F., {Garc{\'\i}a-Burillo}, S., {Audibert}, A., {et~al.} 2019, \aap, 623, A79

\bibitem[{{Combes, F.} {et~al.}(2014){Combes, F.}, {García-Burillo, S.}, {Casasola, V.}, {Hunt, L. K.}, {Krips, M.}, {Baker, A. J.}, {Boone, F.}, {Eckart, A.}, {Marquez, I.}, {Neri, R.}, {Schinnerer, E.}, \& {Tacconi, L. J.}}]{Combes_2014}
{Combes, F.}, {García-Burillo, S.}, {Casasola, V.}, {et~al.} 2014, A\&A, 565, A97

\bibitem[{Comrie {et~al.}(2018)Comrie, Wang, Hsu, Moraghan, Harris, Pang, Pińska, Chiang, Simmonds, Chang, Hwang, Jan, \& Lin}]{CARTA}
Comrie, A., Wang, K.-S., Hsu, S.-C., {et~al.} 2018, {CARTA: The Cube Analysis and Rendering Tool for Astronomy}

\bibitem[{{Condon}(1992)}]{Condon_1992}
{Condon}, J.~J. 1992, \araa, 30, 575

\bibitem[{{Curran} {et~al.}(1998){Curran}, {Johansson}, {Rydbeck}, \& {Booth}}]{Curran_1998}
{Curran}, S.~J., {Johansson}, L.~E.~B., {Rydbeck}, G., \& {Booth}, R.~S. 1998, \aap, 338, 863

\bibitem[{{Di Teodoro} \& {Fraternali}(2015)}]{BBAROLO_2015}
{Di Teodoro}, E.~M. \& {Fraternali}, F. 2015, \mnras, 451, 3021

\bibitem[{{Dorodnitsyn} {et~al.}(2011){Dorodnitsyn}, {Bisnovatyi-Kogan}, \& {Kallman}}]{Dorodnitsyn_2011}
{Dorodnitsyn}, A., {Bisnovatyi-Kogan}, G.~S., \& {Kallman}, T. 2011, \apj, 741, 29

\bibitem[{{Draine}(2011)}]{Draine_2011}
{Draine}, B.~T. 2011, {Physics of the Interstellar and Intergalactic Medium}

\bibitem[{Elmouttie {et~al.}(1998)Elmouttie, Haynes, Jones, Sadler, \& Ehle}]{Elmouttie_1998}
Elmouttie, M., Haynes, R.~F., Jones, K.~L., Sadler, E.~M., \& Ehle, M. 1998, Monthly Notices of the Royal Astronomical Society, 297, 1202

\bibitem[{{Emmering} {et~al.}(1992){Emmering}, {Blandford}, \& {Shlosman}}]{Emmering_1992}
{Emmering}, R.~T., {Blandford}, R.~D., \& {Shlosman}, I. 1992, \apj, 385, 460

\bibitem[{Esquej {et~al.}(2013)Esquej, Alonso-Herrero, González-Martín, Hönig, Hernán-Caballero, Roche, Ramos~Almeida, Mason, Díaz-Santos, Levenson, Aretxaga, Rodríguez~Espinosa, \& Packham}]{Esquej_2014}
Esquej, P., Alonso-Herrero, A., González-Martín, O., {et~al.} 2013, The Astrophysical Journal, 780, 86

\bibitem[{{Fischer} {et~al.}(2013){Fischer}, {Crenshaw}, {Kraemer}, \& {Schmitt}}]{Fischer_2013}
{Fischer}, T.~C., {Crenshaw}, D.~M., {Kraemer}, S.~B., \& {Schmitt}, H.~R. 2013, \apjs, 209, 1

\bibitem[{{Fonseca-Faria} {et~al.}(2021){Fonseca-Faria}, {Rodr{\'\i}guez-Ardila}, {Contini}, \& {Reynaldi}}]{Fonseca_2021}
{Fonseca-Faria}, M.~A., {Rodr{\'\i}guez-Ardila}, A., {Contini}, M., \& {Reynaldi}, V. 2021, \mnras, 506, 3831

\bibitem[{{Freeman} {et~al.}(1977){Freeman}, {Karlsson}, {Lynga}, {Burrell}, {van Woerden}, {Goss}, \& {Mebold}}]{Freeman_1977}
{Freeman}, K.~C., {Karlsson}, B., {Lynga}, G., {et~al.} 1977, \aap, 55, 445

\bibitem[{Gallimore {et~al.}(2004)Gallimore, Baum, \& O’Dea}]{Gallimore_2004}
Gallimore, J.~F., Baum, S.~A., \& O’Dea, C.~P. 2004, The Astrophysical Journal, 613, 794

\bibitem[{{G{\'a}mez Rosas} {et~al.}(2025){G{\'a}mez Rosas}, {van der Werf}, {Gallimore}, {Impellizzeri}, {Jaffe}, {Garc{\'\i}a-Burillo}, {Aalto}, {Burtscher}, {Casasola}, {Combes}, {Henkel}, {M{\'a}rquez}, {Mart{\'\i}n}, {Ramos Almeida}, {Viti}, \& {Fuente}}]{Rosas_2025}
{G{\'a}mez Rosas}, V., {van der Werf}, P., {Gallimore}, J.~F., {et~al.} 2025, \aap, 699, A187

\bibitem[{{Garc{\'\i}a-Burillo} {et~al.}(2016){Garc{\'\i}a-Burillo}, {Combes}, {Ramos Almeida}, {Usero}, {Krips}, {Alonso-Herrero}, {Aalto}, {Casasola}, {Hunt}, {Mart{\'\i}n}, {Viti}, {Colina}, {Costagliola}, {Eckart}, {Fuente}, {Henkel}, {M{\'a}rquez}, {Neri}, {Schinnerer}, {Tacconi}, \& {van der Werf}}]{Burillo_2016}
{Garc{\'\i}a-Burillo}, S., {Combes}, F., {Ramos Almeida}, C., {et~al.} 2016, \apjl, 823, L12

\bibitem[{{Greenhill} {et~al.}(2003){Greenhill}, {Booth}, {Ellingsen}, {Herrnstein}, {Jauncey}, {McCulloch}, {Moran}, {Norris}, {Reynolds}, \& {Tzioumis}}]{Greenhill_2003}
{Greenhill}, L.~J., {Booth}, R.~S., {Ellingsen}, S.~P., {et~al.} 2003, \apj, 590, 162

\bibitem[{{Haardt} \& {Maraschi}(1991)}]{Haardt_1991}
{Haardt}, F. \& {Maraschi}, L. 1991, \apjl, 380, L51

\bibitem[{{Harrison} \& {Ramos Almeida}(2024)}]{Harrison_2024}
{Harrison}, C.~M. \& {Ramos Almeida}, C. 2024, Galaxies, 12, 17

\bibitem[{{Hicks} {et~al.}(2009){Hicks}, {Davies}, {Malkan}, {Genzel}, {Tacconi}, {M{\"u}ller S{\'a}nchez}, \& {Sternberg}}]{Hicks_2009}
{Hicks}, E.~K.~S., {Davies}, R.~I., {Malkan}, M.~A., {et~al.} 2009, \apj, 696, 448

\bibitem[{{Hildebrand}(1983)}]{Hildebrand_1983}
{Hildebrand}, R.~H. 1983, \qjras, 24, 267

\bibitem[{{H{\"o}nig} \& {Kishimoto}(2017)}]{Honig_2017}
{H{\"o}nig}, S.~F. \& {Kishimoto}, M. 2017, \apjl, 838, L20

\bibitem[{{Impellizzeri} {et~al.}(2019){Impellizzeri}, {Gallimore}, {Baum}, {Elitzur}, {Davies}, {Lutz}, {Maiolino}, {Marconi}, {Nikutta}, {O'Dea}, \& {Sani}}]{Impellizzeri_2019}
{Impellizzeri}, C.~M.~V., {Gallimore}, J.~F., {Baum}, S.~A., {et~al.} 2019, \apjl, 884, L28

\bibitem[{Inoue {et~al.}(2020)Inoue, Khangulyan, \& Doi}]{Inoue_2020}
Inoue, Y., Khangulyan, D., \& Doi, A. 2020, The Astrophysical Journal Letters, 891, L33

\bibitem[{{Isbell} {et~al.}(2022){Isbell}, {Meisenheimer}, {Pott}, {Stalevski}, {Tristram}, {Sanchez-Bermudez}, {Hofmann}, {G{\'a}mez Rosas}, {Jaffe}, {Burtscher}, {Leftley}, {Petrov}, {Lopez}, {Henning}, {Weigelt}, {Allouche}, {Berio}, {Bettonvil}, {Cruzalebes}, {Dominik}, {Heininger}, {Hogerheijde}, {Lagarde}, {Lehmitz}, {Matter}, {Meilland}, {Millour}, {Robbe-Dubois}, {Schertl}, {van Boekel}, {Varga}, \& {Woillez}}]{Isbell_2022}
{Isbell}, J.~W., {Meisenheimer}, K., {Pott}, J.~U., {et~al.} 2022, \aap, 663, A35

\bibitem[{{Isbell} {et~al.}(2023){Isbell}, {Pott}, {Meisenheimer}, {Stalevski}, {Tristram}, {Leftley}, {Asmus}, {Weigelt}, {G{\'a}mez Rosas}, {Petrov}, {Jaffe}, {Hofmann}, {Henning}, \& {Lopez}}]{Isbell_2023}
{Isbell}, J.~W., {Pott}, J.~U., {Meisenheimer}, K., {et~al.} 2023, \aap, 678, A136

\bibitem[{Izumi {et~al.}(2018)Izumi, Wada, Fukushige, Hamamura, \& Kohno}]{Izumi_2018}
Izumi, T., Wada, K., Fukushige, R., Hamamura, S., \& Kohno, K. 2018, The Astrophysical Journal, 867, 48

\bibitem[{Izumi {et~al.}(2023)Izumi, Wada, Imanishi, Nakanishi, Kohno, Kudoh, Kawamuro, Baba, Matsumoto, Fujita, \& Tristram}]{Izumi_2023}
Izumi, T., Wada, K., Imanishi, M., {et~al.} 2023, Science, 382, 554

\bibitem[{{Kaiser}(2006)}]{Kaiser_2006}
{Kaiser}, C.~R. 2006, \mnras, 367, 1083

\bibitem[{Kakkad {et~al.}(2023)Kakkad, Stalevski, Kishimoto, Knežević, Asmus, \& Vogt}]{Kakkad_2023}
Kakkad, D., Stalevski, M., Kishimoto, M., {et~al.} 2023, Monthly Notices of the Royal Astronomical Society, 519, 5324

\bibitem[{{Kawamuro} {et~al.}(2022){Kawamuro}, {Ricci}, {Imanishi}, {Mushotzky}, {Izumi}, {Ricci}, {Bauer}, {Koss}, {Trakhtenbrot}, {Ichikawa}, {Rojas}, {Smith}, {Shimizu}, {Oh}, {den Brok}, {Baba}, {Balokovi{\'c}}, {Chang}, {Kakkad}, {Pfeifle}, {Privon}, {Temple}, {Ueda}, {Harrison}, {Powell}, {Stern}, {Urry}, \& {Sanders}}]{Kawamuro_2022}
{Kawamuro}, T., {Ricci}, C., {Imanishi}, M., {et~al.} 2022, \apj, 938, 87

\bibitem[{{Krolik} \& {Begelman}(1986)}]{Krolik_1986}
{Krolik}, J.~H. \& {Begelman}, M.~C. 1986, \apjl, 308, L55

\bibitem[{{Krolik} \& {Begelman}(1988)}]{Krolik_1988}
{Krolik}, J.~H. \& {Begelman}, M.~C. 1988, \apj, 329, 702

\bibitem[{{Maccagni, F. M.} {et~al.}(2021){Maccagni, F. M.}, {Serra, P.}, {Gaspari, M.}, {Kleiner, D.}, {Morokuma-Matsui, K.}, {Oosterloo, T. A.}, {Onodera, M.}, {Kamphuis, P.}, {Loi, F.}, {Thorat, K.}, {Ramatsoku, M.}, {Smirnov, O.}, \& {White, S. V.}}]{Maccagni_2021}
{Maccagni, F. M.}, {Serra, P.}, {Gaspari, M.}, {et~al.} 2021, A\&A, 656, A45

\bibitem[{{Marconi} {et~al.}(1994){Marconi}, {Moorwood}, {Origlia}, \& {Oliva}}]{Marconi_1994}
{Marconi}, A., {Moorwood}, A.~F.~M., {Origlia}, L., \& {Oliva}, E. 1994, The Messenger, 78, 20

\bibitem[{{McCallum} {et~al.}(2009){McCallum}, {Ellingsen}, {Lovell}, {Phillips}, \& {Reynolds}}]{McCallum_2009}
{McCallum}, J.~N., {Ellingsen}, S.~P., {Lovell}, J.~E.~J., {Phillips}, C.~J., \& {Reynolds}, J.~E. 2009, \mnras, 392, 1339

\bibitem[{{Miyamoto} \& {Nagai}(1975)}]{Miyamoto_1975}
{Miyamoto}, M. \& {Nagai}, R. 1975, \pasj, 27, 533

\bibitem[{{Moorwood} {et~al.}(1996){Moorwood}, {Lutz}, {Oliva}, {Marconi}, {Netzer}, {Genzel}, {Sturm}, \& {de Graauw}}]{Moorwood_1996}
{Moorwood}, A.~F.~M., {Lutz}, D., {Oliva}, E., {et~al.} 1996, \aap, 315, L109

\bibitem[{{Mueller Sánchez, F.} {et~al.}(2006){Mueller Sánchez, F.}, {Davies, R. I.}, {Eisenhauer, F.}, {Tacconi, L. J.}, {Genzel, R.}, \& {Sternberg, A.}}]{Mueller_2006}
{Mueller Sánchez, F.}, {Davies, R. I.}, {Eisenhauer, F.}, {et~al.} 2006, A\&A, 454, 481

\bibitem[{Mutie {et~al.}(2025)Mutie, del Palacio, Beswick, Williams-Baldwin, Gallimore, Gallagher, Aalto, \& Baki}]{Mutie_2025}
Mutie, I.~M., del Palacio, S., Beswick, R.~J., {et~al.} 2025, Monthly Notices of the Royal Astronomical Society, staf524

\bibitem[{{Oliva} {et~al.}(1994){Oliva}, {Salvati}, {Moorwood}, \& {Marconi}}]{Oliva_1994}
{Oliva}, E., {Salvati}, M., {Moorwood}, A.~F.~M., \& {Marconi}, A. 1994, \aap, 288, 457

\bibitem[{{Palit, B.} {et~al.}(2024){Palit, B.}, {Różańska, A.}, {Petrucci, P. O.}, {Gronkiewicz, D.}, {Barnier, S.}, {Bianchi, S.}, {Ballantyne, D. R.}, {Gianolli, V. E.}, {Middei, R.}, {Belmont, R.}, \& {Ursini, F.}}]{Palit_2024}
{Palit, B.}, {Różańska, A.}, {Petrucci, P. O.}, {et~al.} 2024, A\&A, 690, A308

\bibitem[{{Panessa} {et~al.}(2019){Panessa}, {Baldi}, {Laor}, {Padovani}, {Behar}, \& {McHardy}}]{Panessa_2019}
{Panessa}, F., {Baldi}, R.~D., {Laor}, A., {et~al.} 2019, Nature Astronomy, 3, 387

\bibitem[{{Parma} {et~al.}(1999){Parma}, {Murgia}, {Morganti}, {Capetti}, {de Ruiter}, \& {Fanti}}]{Parma_1999}
{Parma}, P., {Murgia}, M., {Morganti}, R., {et~al.} 1999, \aap, 344, 7

\bibitem[{{Patrikeev} {et~al.}(2006){Patrikeev}, {Fletcher}, {Stepanov}, {Beck}, {Berkhuijsen}, {Frick}, \& {Horellou}}]{Patrikeev_2006}
{Patrikeev}, I., {Fletcher}, A., {Stepanov}, R., {et~al.} 2006, \aap, 458, 441

\bibitem[{{Pier} \& {Krolik}(1992)}]{Pier_1992}
{Pier}, E.~A. \& {Krolik}, J.~H. 1992, \apjl, 399, L23

\bibitem[{{Pogge} \& {Martini}(2002)}]{Pogge_2002}
{Pogge}, R.~W. \& {Martini}, P. 2002, \apj, 569, 624

\bibitem[{{Prieto} {et~al.}(2005){Prieto}, {Maciejewski}, \& {Reunanen}}]{Prieto_2005}
{Prieto}, M.~A., {Maciejewski}, W., \& {Reunanen}, J. 2005, \aj, 130, 1472

\bibitem[{Ricci {et~al.}(2023)Ricci, Chang, Kawamuro, Privon, Mushotzky, Trakhtenbrot, Laor, Koss, Smith, Gupta, Dimopoulos, Aalto, \& Ros}]{Ricci_2023}
Ricci, C., Chang, C.-S., Kawamuro, T., {et~al.} 2023, The Astrophysical Journal Letters, 952, L28

\bibitem[{Rivers {et~al.}(2011)Rivers, Markowitz, \& Rothschild}]{Rivers_2011}
Rivers, E., Markowitz, A., \& Rothschild, R. 2011, The Astrophysical Journal Supplement Series, 193, 3

\bibitem[{{Roberts} {et~al.}(1979){Roberts}, {Huntley}, \& {van Albada}}]{Roberts_1979}
{Roberts}, Jr., W.~W., {Huntley}, J.~M., \& {van Albada}, G.~D. 1979, \apj, 233, 67

\bibitem[{{Rybicki} \& {Lightman}(1979)}]{Rybicki_1979}
{Rybicki}, G.~B. \& {Lightman}, A.~P. 1979, {Radiative processes in astrophysics}

\bibitem[{Schartmann {et~al.}(2017)Schartmann, Mould, Wada, Burkert, Durré, Behrendt, Davies, \& Burtscher}]{Schartmann_2017}
Schartmann, M., Mould, J., Wada, K., {et~al.} 2017, Monthly Notices of the Royal Astronomical Society, 473, 953

\bibitem[{Schnorr~Müller {et~al.}(2011)Schnorr~Müller, Storchi-Bergmann, Riffel, Ferrari, Steiner, Axon, \& Robinson}]{Muller_2011}
Schnorr~Müller, A., Storchi-Bergmann, T., Riffel, R.~A., {et~al.} 2011, Monthly Notices of the Royal Astronomical Society, 413, 149

\bibitem[{{Sch{\"o}ier} {et~al.}(2005){Sch{\"o}ier}, {van der Tak}, {van Dishoeck}, \& {Black}}]{LAMDA_2005}
{Sch{\"o}ier}, F.~L., {van der Tak}, F.~F.~S., {van Dishoeck}, E.~F., \& {Black}, J.~H. 2005, \aap, 432, 369

\bibitem[{{Shirley}(2015)}]{Shirley_2015}
{Shirley}, Y.~L. 2015, \pasp, 127, 299

\bibitem[{Shu {et~al.}(2011)Shu, Yaqoob, \& Wang}]{Shu_2011}
Shu, X.~W., Yaqoob, T., \& Wang, J.~X. 2011, The Astrophysical Journal, 738, 147

\bibitem[{{Stalevski} {et~al.}(2017){Stalevski}, {Asmus}, \& {Tristram}}]{Stalevski_2017}
{Stalevski}, M., {Asmus}, D., \& {Tristram}, K. R.~W. 2017, \mnras, 472, 3854

\bibitem[{Stalevski {et~al.}(2019)Stalevski, Tristram, \& Asmus}]{Stalevski_2019}
Stalevski, M., Tristram, K. R.~W., \& Asmus, D. 2019, Monthly Notices of the Royal Astronomical Society, 484, 3334

\bibitem[{{Storchi-Bergmann} {et~al.}(1992){Storchi-Bergmann}, {Mulchaey}, \& {Wilson}}]{Storchi_1992}
{Storchi-Bergmann}, T., {Mulchaey}, J.~S., \& {Wilson}, A.~S. 1992, \apjl, 395, L73

\bibitem[{Storchi-Bergmann \& Schnorr-Müller(2019)}]{Storchi_Bergmann_2019}
Storchi-Bergmann, T. \& Schnorr-Müller, A. 2019, Nature Astronomy, 3, 48–61

\bibitem[{{Sz{\H{u}}cs} {et~al.}(2016){Sz{\H{u}}cs}, {Glover}, \& {Klessen}}]{Laszlo_2016}
{Sz{\H{u}}cs}, L., {Glover}, S. C.~O., \& {Klessen}, R.~S. 2016, \mnras, 460, 82

\bibitem[{{The CASA Team} {et~al.}(2022){The CASA Team}, Bean, Bhatnagar, Castro, Meyer, Emonts, Garcia, Garwood, Golap, Villalba, Harris, Hayashi, Hoskins, Hsieh, Jagannathan, Kawasaki, Keimpema, Kettenis, Lopez, Marvil, Masters, McNichols, Mehringer, Miel, Moellenbrock, Montesino, Nakazato, Ott, Petry, Pokorny, Raba, Rau, Schiebel, Schweighart, Sekhar, Shimada, Small, Steeb, Sugimoto, Suoranta, Tsutsumi, van Bemmel, Verkouter, Wells, Xiong, Szomoru, Griffith, Glendenning, \& Kern}]{CASA}
{The CASA Team}, Bean, B., Bhatnagar, S., {et~al.} 2022, Publications of the Astronomical Society of the Pacific, 134, 114501

\bibitem[{{Toomre}(1964)}]{Toomre_1964}
{Toomre}, A. 1964, \apj, 139, 1217

\bibitem[{{Tristram} {et~al.}(2014){Tristram}, {Burtscher}, {Jaffe}, {Meisenheimer}, {H{\"o}nig}, {Kishimoto}, {Schartmann}, \& {Weigelt}}]{Tristram_2014}
{Tristram}, K. R.~W., {Burtscher}, L., {Jaffe}, W., {et~al.} 2014, \aap, 563, A82

\bibitem[{{Tristram} {et~al.}(2022){Tristram}, {Impellizzeri}, {Zhang}, {Villard}, {Henkel}, {Viti}, {Burtscher}, {Combes}, {Garc{\'\i}a-Burillo}, {Mart{\'\i}n}, {Meisenheimer}, \& {van der Werf}}]{Tristram_2022}
{Tristram}, K. R.~W., {Impellizzeri}, C.~M.~V., {Zhang}, Z.-Y., {et~al.} 2022, \aap, 664, A142

\bibitem[{Tristram {et~al.}(2007)Tristram, Meisenheimer, Jaffe, Schartmann, Rix, Leinert, Morel, Wittkowski, Röttgering, Perrin, Lopez, Raban, Cotton, Graser, Paresce, \& Henning}]{Tristram_2007}
Tristram, K. R.~W., Meisenheimer, K., Jaffe, W., {et~al.} 2007, Astronomy \& Astrophysics, 474, 837–850

\bibitem[{{Urry} \& {Padovani}(1995)}]{Urry_1995}
{Urry}, C.~M. \& {Padovani}, P. 1995, \pasp, 107, 803

\bibitem[{Vall\'{e}e(2020)}]{Vallee_2020}
Vall\'{e}e, J.~P. 2020, New Astronomy, 76, 101337

\bibitem[{{van de Ven} \& {Fathi}(2010)}]{VandeVen_2010}
{van de Ven}, G. \& {Fathi}, K. 2010, \apj, 723, 767

\bibitem[{{van der Tak} {et~al.}(2007){van der Tak}, {Black}, {Sch{\"o}ier}, {Jansen}, \& {van Dishoeck}}]{Radex_2007}
{van der Tak}, F.~F.~S., {Black}, J.~H., {Sch{\"o}ier}, F.~L., {Jansen}, D.~J., \& {van Dishoeck}, E.~F. 2007, \aap, 468, 627

\bibitem[{{Viti} {et~al.}(2014){Viti}, {Garc{\'\i}a-Burillo}, {Fuente}, {Hunt}, {Usero}, {Henkel}, {Eckart}, {Martin}, {Spaans}, {Muller}, {Combes}, {Krips}, {Schinnerer}, {Casasola}, {Costagliola}, {Marquez}, {Planesas}, {van der Werf}, {Aalto}, {Baker}, {Boone}, \& {Tacconi}}]{Viti_2014}
{Viti}, S., {Garc{\'\i}a-Burillo}, S., {Fuente}, A., {et~al.} 2014, \aap, 570, A28

\bibitem[{{Vollmer} {et~al.}(2018){Vollmer}, {Schartmann}, {Burtscher}, {Marin}, {H{\"o}nig}, {Davies}, \& {Goosmann}}]{Vollmer_2018}
{Vollmer}, B., {Schartmann}, M., {Burtscher}, L., {et~al.} 2018, \aap, 615, A164

\bibitem[{Wada(2012)}]{Wada_2012}
Wada, K. 2012, The Astrophysical Journal, 758, 66

\bibitem[{{Wada} \& {Norman}(2002)}]{Wada_2002}
{Wada}, K. \& {Norman}, C.~A. 2002, \apjl, 566, L21

\bibitem[{Wada {et~al.}(2016)Wada, Schartmann, \& Meijerink}]{Wada_2016}
Wada, K., Schartmann, M., \& Meijerink, R. 2016, The Astrophysical Journal Letters, 828, L19

\bibitem[{{Wada} {et~al.}(2018){Wada}, {Yonekura}, \& {Nagao}}]{Wada_2018}
{Wada}, K., {Yonekura}, K., \& {Nagao}, T. 2018, \apj, 867, 49

\bibitem[{Warmels {et~al.}(2018)Warmels, Biggs, Cortes, Dent, Di~Francesco, Fomalont, Hales, Kameno, Mason, Philips, Remijan, Saini, Stoehr, Vila~Vilaro, \& Villard}]{ALMA_handbook}
Warmels, R., Biggs, A., Cortes, P., {et~al.} 2018, ALMA Technical Handbook, ALMA Doc. 6.3, ver. 1.0

\bibitem[{{Yang} {et~al.}(2010){Yang}, {Stancil}, {Balakrishnan}, \& {Forrey}}]{Lamda_CO_collrates}
{Yang}, B., {Stancil}, P.~C., {Balakrishnan}, N., \& {Forrey}, R.~C. 2010, \apj, 718, 1062

\bibitem[{Yang {et~al.}(2009)Yang, Wilson, Matt, Terashima, \& Greenhill}]{Yang_2009}
Yang, Y., Wilson, A.~S., Matt, G., Terashima, Y., \& Greenhill, L.~J. 2009, The Astrophysical Journal, 691, 131

\bibitem[{{Zschaechner} {et~al.}(2016){Zschaechner}, {Walter}, {Bolatto}, {Farina}, {Kruijssen}, {Leroy}, {Meier}, {Ott}, \& {Veilleux}}]{Zschaechner_2016}
{Zschaechner}, L.~K., {Walter}, F., {Bolatto}, A., {et~al.} 2016, \apj, 832, 142

\end{thebibliography}

\begin{appendix}
\onecolumn
\section{Observational details and results}\label{sec:app_obs}

\begin{table*}[h]
    \centering
    \caption{Observation and data reduction details for the auxiliary data}

    \begin{tabular}{c | c c c c c c} 
        \hline
        \hline
         & B3 LSB & B3 USB & B4 LSB & B4 USB & B5 & B9\\
        \hline
        $\nu_{\mathrm{central}}$ [GHz] & 85.963 & 98.732 & 135.374 & 146.834 & 178.009 & 664.599\\
        $\lambda_{\mathrm{central}}$ [mm] & 3.49 & 3.04 & 2.21 & 2.04 & 1.68 & 0.45\\
        Beam [mas$^2$] & $65.2\times48.2$ & $57.8\times42.1$ & $61.0\times41.5$ & $53.8\times38.1$ & $61.3\times46.6$ & $74.5\times55.9$\\
        Beam [pc$^2$] & $1.3\times1.0$ & $1.2\times0.8$ & $1.2\times0.8$ & $1.1\times0.8$ & $1.2\times0.9$ & $1.5\times1.1$\\
        P.A. [$^{\circ}$] & -15.6 & -15.3 & -0.9 & -5.6 & 40.8 & 71.3\\
        MRS ["] & 1.132 & 1.132 & 0.702 & 0.675 & 1.107 & 1.440\\
        Int. time ['] & 49 & 49 & 22 & 22 & 118 & 126\\
        Sens. [$\mu$Jy/bm] & 219 & 211 & 38.6 & 38.7 & 32.0 & 330\\
        Obs. date & 2017-10-08 & 2017-10-08 & 2017-10-15 & 2017-10-15 & 2021-08-20 & 2021-07-01\\
        Pipeline version & 40896 & 40896  &  40896 &  40896 & 2020.1.0.40 & 2021.2.0.128\\
        Self-cal iter.: & 0 & 0 & 0 & 0 & 0 & 0\\
        \hline
    \end{tabular}
    \tablefoot{The cited beam sizes represent the beam sizes of the respective continuum maps imaged using Briggs weighting (robust parameter $=0.5$). The maximal recoverable scales (MRS) were derived using Eq. (3.27) in \citet{ALMA_handbook}.} \label{table:measurement_sets_auxiliary}
\end{table*}

\begin{table*}[h]
    \centering
    \caption{Continuum flux density measurements used for the SED in Fig. \ref{fig:SED}}

    \begin{tabular}{c c c c c} 
        \hline
        \hline
        $\nu\,\mathrm{[GHz]}$ & $\lambda\,\mathrm{[mm]}$ & Position & $\mathrm{Beam/Aperture\,size}\,\mathrm{[mas}^2\mathrm{]}$ & $F_{\nu}\,\mathrm[mJy]$\\
        \hline
        664.60 & 0.451 & Nucleus & $74.5\times65.2$ & $16.74\pm1.70$\\
               &       &         & $300\times300$   & $82.93\pm8.58$\\
               &       & Phoenix & $200\times200$   & $17.64\pm2.29$\\
        349.56 & 0.858 & Nucleus & $74.5\times65.2$ & $21.96\pm2.20$\\
               &       &         & $300\times300$   & $32.65\pm3.29$\\
               &       & Phoenix & $200\times200$   & $0.98\pm0.28$ \\
        258.82 & 1.158 & Nucleus & $74.5\times65.2$ & $20.97\pm2.10$\\
               &       &         & $300\times300$   & $32.65\pm3.29$\\
               &       & Phoenix & $200\times200$   & $0.54\pm0.16$ \\
        178.01 & 1.684 & Nucleus & $74.5\times65.2$ & $16.87\pm1.69$\\
               &       &         & $300\times300$   & $26.82\pm2.70$\\
        146.83 & 2.042 & Nucleus & $74.5\times65.2$ & $18.35\pm1.84$\\
               &       &         & $300\times300$   & $27.67\pm2.80$\\
        135.37 & 2.215 & Nucleus & $74.5\times65.2$ & $19.11\pm1.91$\\
               &       &         & $300\times300$   & $28.44\pm2.87$\\
        98.73  & 3.036 & Nucleus & $74.5\times65.2$ & $16.58\pm1.67$\\
               &       &         & $300\times300$   & $28.67\pm3.27$\\
        85.96  & 3.487 & Nucleus & $74.5\times65.2$ & $19.87\pm2.00$\\
               &       &         & $300\times300$   & $33.47\pm3.78$\\
        8.64   & 34.7  & Nucleus & $\left(0.9\times10^3\right)\times\left(0.8\times10^3\right)$   & $50\pm5$      \\
        4.80   & 62.5  & Nucleus & $\left(1.4\times10^3\right)\times\left(1.3\times10^3\right)$   & $50\pm5$      \\
        2.37   & 126   & Nucleus & $\left(3.4\times10^3\right)\times\left(3.1\times10^3\right)$   & $70\pm7$      \\
        1.37   & 219  & Nucleus  & $\left(6.0\times10^3\right)\times\left(5.4\times10^3\right)$   & $120\pm12$    \\
        \hline
    \end{tabular}
    \tablefoot{Columns list the central frequency ($\nu$) and wavelength ($\lambda$), the position of the measurement (either the Nucleus at $\alpha=14^{\mathrm{h}}13^{\mathrm{m}}09.9473^{\mathrm{s}}$, $\delta=-65^{\circ}20'21.023"$ or the Phoenix feature at $\alpha=14^{\mathrm{h}}13^{\mathrm{m}}09.9850^{\mathrm{s}}$, $\delta=-65^{\circ}20'21.271"$; ICRS), the beam size (for nucleus measurements) or aperture diameter (for Phoenix), and the flux density. The four lowest frequency measurements were made by the ATCA radio telescope and are taken from \citet{Elmouttie_1998}.} \label{table:SED_data}
\end{table*}

\begin{figure*}[h]
    \centering
    \includegraphics[width=0.98\linewidth]{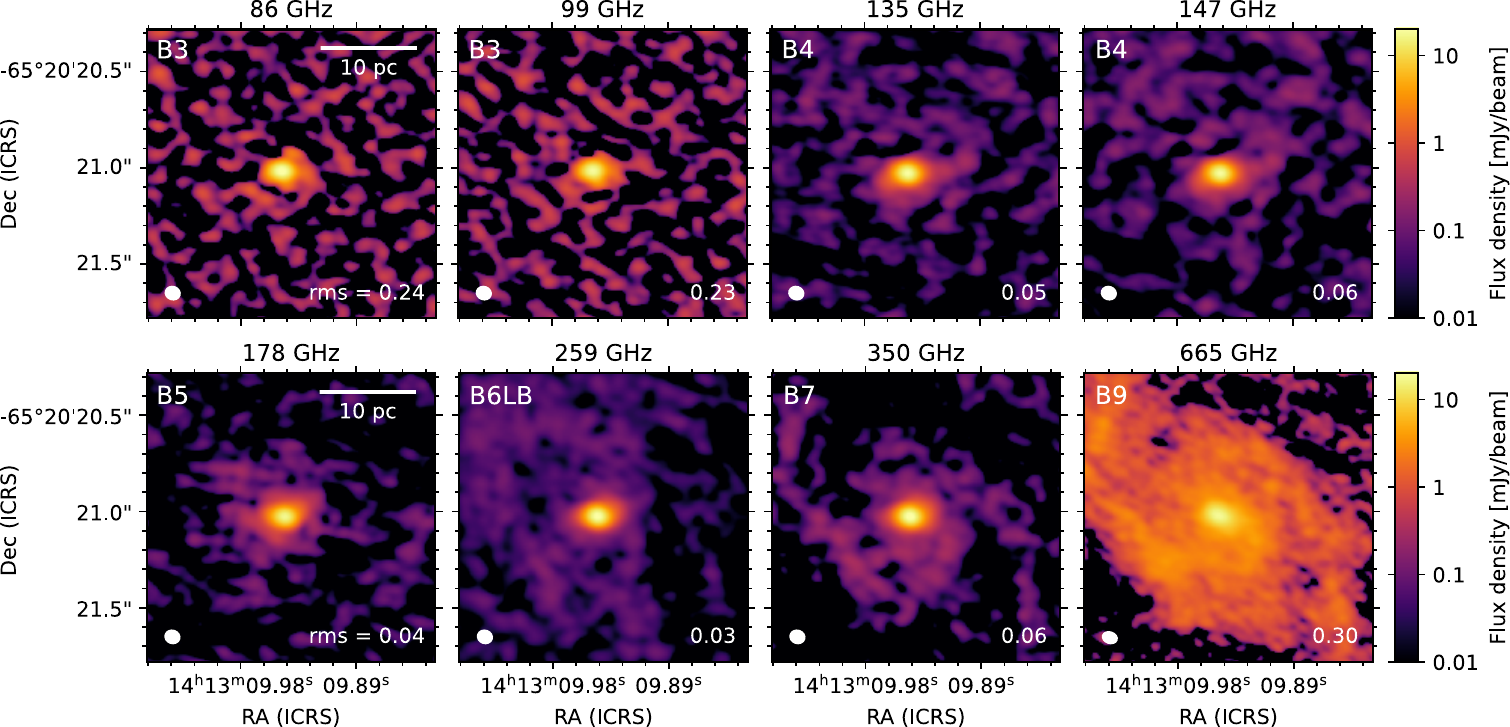}
    \caption{Continuum maps of the Circinus nucleus at the eight frequencies used for the SED decomposition presented in Sect. \ref{sec:continuum_origin}. All maps were convolved to a beam size of $74.5\times65.2\,\mathrm{mas}^2$. We discern two emission components: free-free emission at the nucleus, which is visible in all bands, and dust emission, which is only visible in band~6, 7 and 9. We note that some extended emission is likely missing in the B6LB map because it has an MRS of 375 mas. This is likely also the reason for any discrepancy between this map and the one presented in Fig. \ref{fig:cont_morphology}. The synthesised beam sizes are drawn in the bottom left as filled ellipses, and the background rms values are shown in the bottom right in units of mJy/beam.}
    \label{fig:cont_SED}
\end{figure*}

\begin{figure*}[h]
    \centering
    \includegraphics[width=0.99\linewidth]{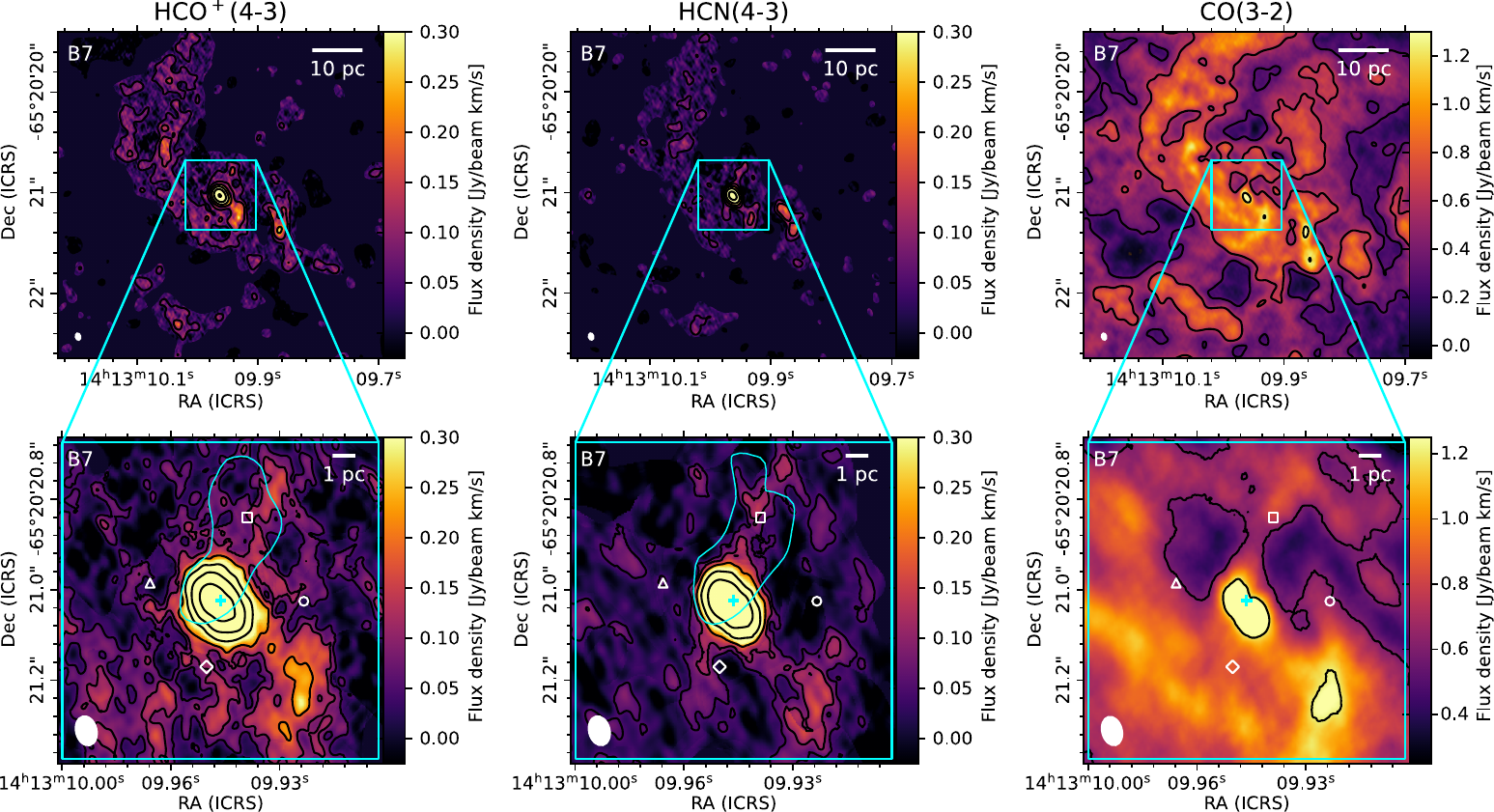}
    \caption{Band 7 HCO$^+$(4-3) ($\nu_{\mathrm{rest}}=356.7\,\mathrm{GHz}$), HCN(4-3) ($\nu_{\mathrm{rest}}=354.5\,\mathrm{GHz}$), and CO(3-2) ($\nu_{\mathrm{rest}}=345.8\,\mathrm{GHz}$) moment zero maps. We observe spiral structure in the CND that connects to the torus in all three transitions, with the bottom panels showing the inner bars of these hooks. Contour levels are drawn at $[1, 2, 4, 8, 16, 32]\,\times\,50\,\mathrm{mJy/beam}\,\mathrm{km\,s}^{-1}$ in the HCO$^+$(4-3) and HCN(4-3) maps and at $[1, 2, 4]\,\times\,300\,\mathrm{mJy/beam}\,\mathrm{km\,s}^{-1}$ in the CO(3-2) maps. Cyan contours in the bottom left and middle panels indicate the position of the blue-shifted emission feature discussed in Sect. \ref{sec:inflow} and represent the $-30\,\mathrm{km\,s^{-1}}$ level in right panels of Fig. \ref{fig:BBAROLO_velfield} for HCO$^+$(3-2) and HCN(3-2), respectively. The synthesised beam sizes are drawn in the bottom left as filled ellipses, and the cyan cross indicates the AGN position. Indicated on the bottom panels are also the four locations where the 1-D kinematic profiles from Fig. \ref{fig:1D_kin_profiles} were extracted using corresponding symbols in white. To boost the S/N in the moment zero maps, we applied a custom masking algorithm for the HCO$^+$(4-3) and HCN(4-3) maps that conservatively masks out empty regions in the image cube. For CO(3-2), we masked out negative values in the image cube to remove artefacts caused by large scale emission that exceeds the maximal recoverable scale. Since this causes an upward bias in the flux density, the resulting CO(3-2) moment zero map is unsuitable for quantitative flux measurements.}
    \label{fig:mol_morphology_B7}
\end{figure*}

\twocolumn

\section{Continuum analysis}\label{sec:app_cont}
\subsection{Possible synchrotron contribution in the mm regime from a jet or the X-ray corona}\label{sec:synchrotron_contribution}
Based on the SED (Fig. \ref{fig:SED}), we find no clear evidence for synchrotron emission, which typically follows a power-law slope of $\alpha = -0.75$ in the optically thin regime and $\alpha = 2.5$ when optically thick \citep{Rybicki_1979}. Synchrotron emission from a compact nuclear jet can, however, produce a flat spectrum at millimetre wavelengths, as is commonly observed in AGN (e.g., \citealt{Kaiser_2006, Kawamuro_2022}). Such emission is characterised by high brightness temperatures, typically $T_{\mathrm{B}} > 10^7\,\mathrm{K}$ \citep{Panessa_2019}. In our highest-resolution map (B6LB; $14.3 \times 12.2$ mas$^2$), the peak flux density is $1.26 \times 10^{-2}\,\mathrm{Jy/beam}$, corresponding to an antenna temperature of $T_{\mathrm{A}}=1.3\times10^3\,\mathrm{K}$. Since the brightness temperature of the resolved emission remains below this peak value, synchrotron radiation is unlikely to dominate there. The central core is unresolved, however, which leaves open the possibility that $T_{\mathrm{B}} \gg T_{\mathrm{A}}$\footnote{The antenna temperature is equal to the brightness temperature only if a source is resolved.}. A relevant analogue where this central core component is resolved is NGC 1068, which is also known to contain a nuclear jet. In NGC 1068, \citet{Gallimore_2004} identified a similar pc-scale free-free emitting region (component “S1”) with $T_{\mathrm{B}}=4\times10^6\,\mathrm{K}$ that originates near the X-ray-irradiated maser disk. In that case, the free-free component dominates over the synchrotron jet in the mm continuum  \citep{Mutie_2025}. Since no jet has been observed in Circinus, we expect that any jet contribution in Circinus is even weaker than in NGC 1068. We therefore interpret the unresolved mm-core emission as arising from free-free processes in the X-ray corona or the maser disk, rather than from a nuclear jet (see also \citealt{Blundell_2007}).

In addition to a nuclear jet, X-ray coronae can also contribute to the synchrotron emission observed in AGN (see e.g. \citet{Inoue_2020} and \citet{Mutie_2025} in the case of NGC 1068). In fact, based on the empirical correlation between X-ray and mm brightness from \citet{Ricci_2023} and the observed X-ray flux of $F_{2-10\,\mathrm{keV}}=[1.3-2.6]\times10^{-11}\,\mathrm{erg}\,\mathrm{cm}^{-2}\,\mathrm{s}^{-1}$ for the Circinus AGN \citep{Beckmann_2006, Shu_2011, Rivers_2011, Yang_2009}, we expect a self-absorbed synchrotron component of $F_{100\,\mathrm{GHz}}=[1.9-33]\,\mathrm{mJy}$ to exist in Circinus that peaks at mm wavelengths. Since we do not observe a peaked synchrotron component in the SED, we conclude that if present, it must be significantly weaker than the free-free component in the unresolved core. We therefore conclude that the nuclear continuum emission is largely dominated by optically thin thermal free-free emission between 86 GHz and 665 GHz and does not contain a significant contribution from synchrotron radiation.

\subsection{Matching the VLBI maser astrometry to that of B6LB}\label{sec:VLBI_astrometry}
The VLBI masers from \citet{McCallum_2009} presented in Sect. \ref{sec:ionised_outflow} have small relative positional uncertainty ($\sim0.1\,\mathrm{mas}$) but lack absolute astrometry. To solve this, we matched the position of the brightest 22 GHz maser from \citet{McCallum_2009} to the astrometry of the brightest 183 GHz maser from S. Venselaar et al. (in prep.), which have almost the same radial velocity ($v_{r,22\,\mathrm{GHz}}=126.5\,\mathrm{km\,s}^{-1}$ and $v_{r,183\,\mathrm{GHz}}=125.4\,\mathrm{km\,s}^{-1}$) and overall position in the disk and can thus be assumed to belong to the same maser cloud. We then matched the astrometry from S. Venselaar et al. (in prep.) to that of B6LB by aligning the $183\,\mathrm{GHz}$ continuum peak to the position of the continuum peak in B6LB. We estimate that the resulting astrometric uncertainty of the maser positions on our B6LB map is $\sim6\,\mathrm{mas}$. From the relative maser positions, we can also derive an estimate for the position of the kinematic centre. To do this, we followed the method from \citet{Greenhill_2003} by taking the average position of the innermost blue- and red-shifted 22 GHz masers. We find that the kinematic centre is located $5.3\pm6.0\,\mathrm{mas}$ to the north-west of the continuum peak, which is an insignificant offset within the relative astrometric uncertainty.

\section{Additional kinematic modelling}\label{sec:app_kin}
\subsection{Major axis PV diagram}\label{sec:PV_diagram}

\begin{figure}[ht]
    \centering
    \includegraphics[width=0.97\linewidth]{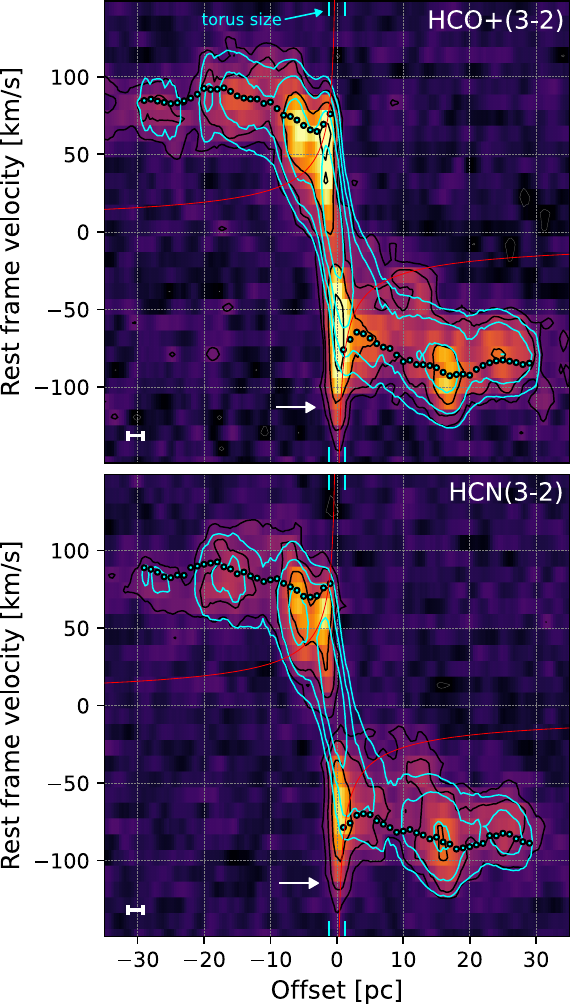}
    \caption{Position-velocity (PV) diagrams along the major axis for HCO$^+$(3-2) (top, PA=$216^{\circ}$) and HCN(3-2) (bottom, PA=$217^{\circ}$) in the B6SB dataset. The colour map and black contours represent the data, and the cyan contours represent the model. Both sets of contours are drawn at $[1,2,4,8]\times1.5\,\mathrm{mJy/beam}$. Overlaid on the PV diagrams are also the fitted line-of-sight velocities $V_{\mathrm{LOS}}=\sin(i)\times V_{\mathrm{rot}}$ using cyan circles and the theoretical Keplerian profile of a central SMBH with a mass of $M_{\bullet}=1.7\times10^6\,\mathrm{M}_{\odot}$ \citep{Greenhill_2003} using a red solid curve. The size of the torus, which is barely resolved in B6SB, is indicated on the edges of each panel using vertical cyan markers. A white arrow is drawn in both panels to indicate the location of a $<-100\,\mathrm{km\,s^{-1}}$ blue-shifted emission component in the data that is not replicated by the model. The synthesised beam sizes are shown in the bottom left as white bars.}
    \label{fig:BBAROLO_PV}
\end{figure}

The major axis PV diagram presented in Fig. \ref{fig:BBAROLO_PV} shows that the $^\mathrm{3D}$BAROLO model is generally able to reproduce the kinematics observed in the CND along the major axis. However, within the central pc, we see that the model is not able to properly reproduce the double-peaked kinematic profile and instead smears into one broad component. This inability is likely due to the inhomogeneous structure of the torus emission, which is unresolved in B6SB. Because $^\mathrm{3D}$BAROLO has no knowledge of spatial emission structure that is smaller than the beam size, it assumes that the inner ring has a homogeneous flux density, which includes both the receding and approaching sides of the torus and portions of the disk that are at systemic velocity. However, we know from the moment zero maps of B6LB (bottom panels in Fig. \ref{fig:mol_morphology}) that this assumption is false and that the regions with a systemic line-of-sight velocity (along the minor axis) are weaker or completely missing due to the central absorption feature, especially in HCN(3-2). Looking at the PV diagrams in Fig. \ref{fig:BBAROLO_PV}, we see that indeed this gap in the data at systemic velocity is stronger in HCN(3-2) than in HCO$^{+}$(3-2). We do not expect this discrepancy between the model and the data to have an effect on the modelling of the surrounding CND. Another discrepancy is the blue wing that we observe towards the torus (indicated with a white arrow in Fig. \ref{fig:BBAROLO_PV}). This blue wing causes an asymmetry in the torus spectrum between both sides of the torus. Its origin remains unclear, but possible reasons may be that we see fast-moving gas clouds that are orbiting the SMBH at the inner edge of the torus or, alternatively, a molecular outflow that is launched near the AGN.

\subsection{Sensitivity of residual velocities to model parameters}\label{sec:improper_modelling}
To test whether the strong residuals described in Sect. \ref{sec:CND_kinematics} are the result of improper modelling of (a part of) the disk, we experimented with fixing different model parameters. Changing $V_{\mathrm{sys}}$ shifts the residuals, causing the blue (red) residuals to be stronger when choosing a higher (lower) $V_{\mathrm{sys}}$. It is impossible to get rid of either the blue or red residuals completely without causing the rest of the disk to show a general residual blue or redshift that implies an improper value of $V_{\mathrm{sys}}$. Forcing the PA to be $190^{\circ}$ instead of $218\pm6^{\circ}$ eliminates the north-western blue and southern red residuals. However, this in turn creates a strong dipole in the residuals along the previously fitted position angle of $218\pm6^{\circ}$ at all radii, indicating that this fails to properly model the dominant rotation of the disk. Changing the inclination also changes the magnitude of the NW blue and S red residuals, with a larger (smaller) inclination angle producing larger (smaller) residuals. However, one needs to force the inclination angle to be $i\leq50^{\circ}$ for the blue and red residuals to become insignificant compared to the other residual velocities in the disk, at which point the expected disk morphology does not match the observed oblateness of the CND. In addition, the inclination would have to differ significantly from the values found by \citetalias{Tristram_2022} and \citet{Curran_1998} ($i=75\pm5^{\circ}$) at the same spatial scale, \citet{Izumi_2018} ($i\approx70\pm5^{\circ}$) at the 100 pc scale and by \citetalias{Izumi_2023} ($i\sim82^{\circ}$) at the pc scale, which we consider unrealistic. Lastly, we experimented with only fitting the approaching or receding sides of the CND with $^\mathrm{3D}$BAROLO. We find that the fit is robust against fitting only one side and that, consequently, the observed blue and red residuals do not change significantly. We therefore conclude that the observed velocity residuals are not the result of improper modelling.

\subsection{Kinematic modelling with non-zero radial velocity}\label{sec:Vrad_free}
\begin{figure}[h]
    \centering
    \includegraphics[width=\linewidth]{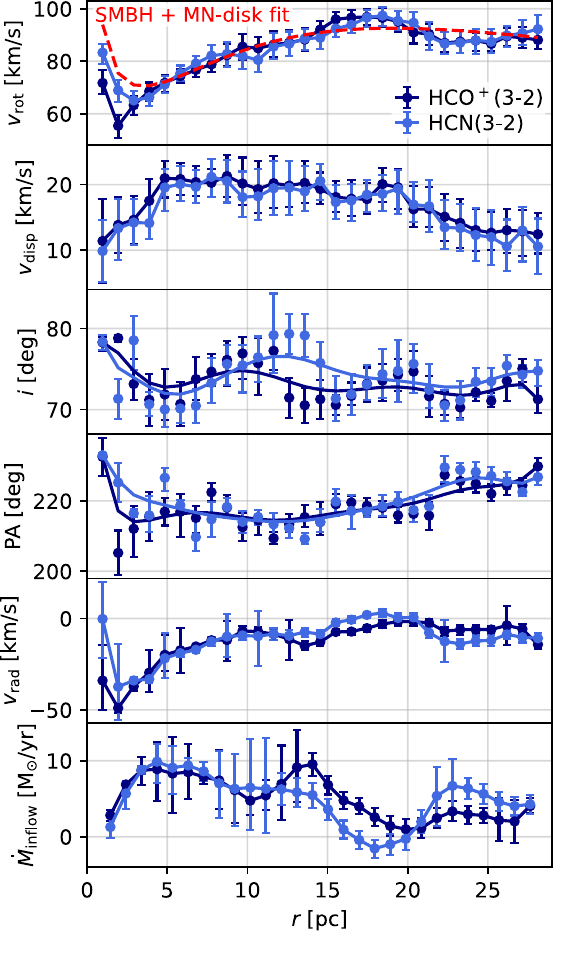}
    \caption{Same as Fig. \ref{fig:BBAROLO_params}, but for $V_{\mathrm{rad}}$ left as a free parameter. The fitted radial $V_{\mathrm{rad}}$ profile is shown in the fifth panel, where we define $V_{\mathrm{rad}}$ such that negative (positive) values imply inflow (outflow). In the bottom panel, we show the radial inflow rate that results from multiplying $V_{\mathrm{rad}}(r)$ by the radial mass distribution $dM_{\mathrm{MN}}(r)/dr$ of the SMBH + Miyamoto-Nagai disk model (red dashed curve in the top panel, which was fitted to the rotation curve for $V_{\mathrm{rad}}=0$, see Sect. \ref{sec:inflow_rate}).}
    \label{fig:BBAROLO_params_Vradfree}
\end{figure}

\begin{figure*}[h]
    \centering
    \includegraphics[width=\linewidth]{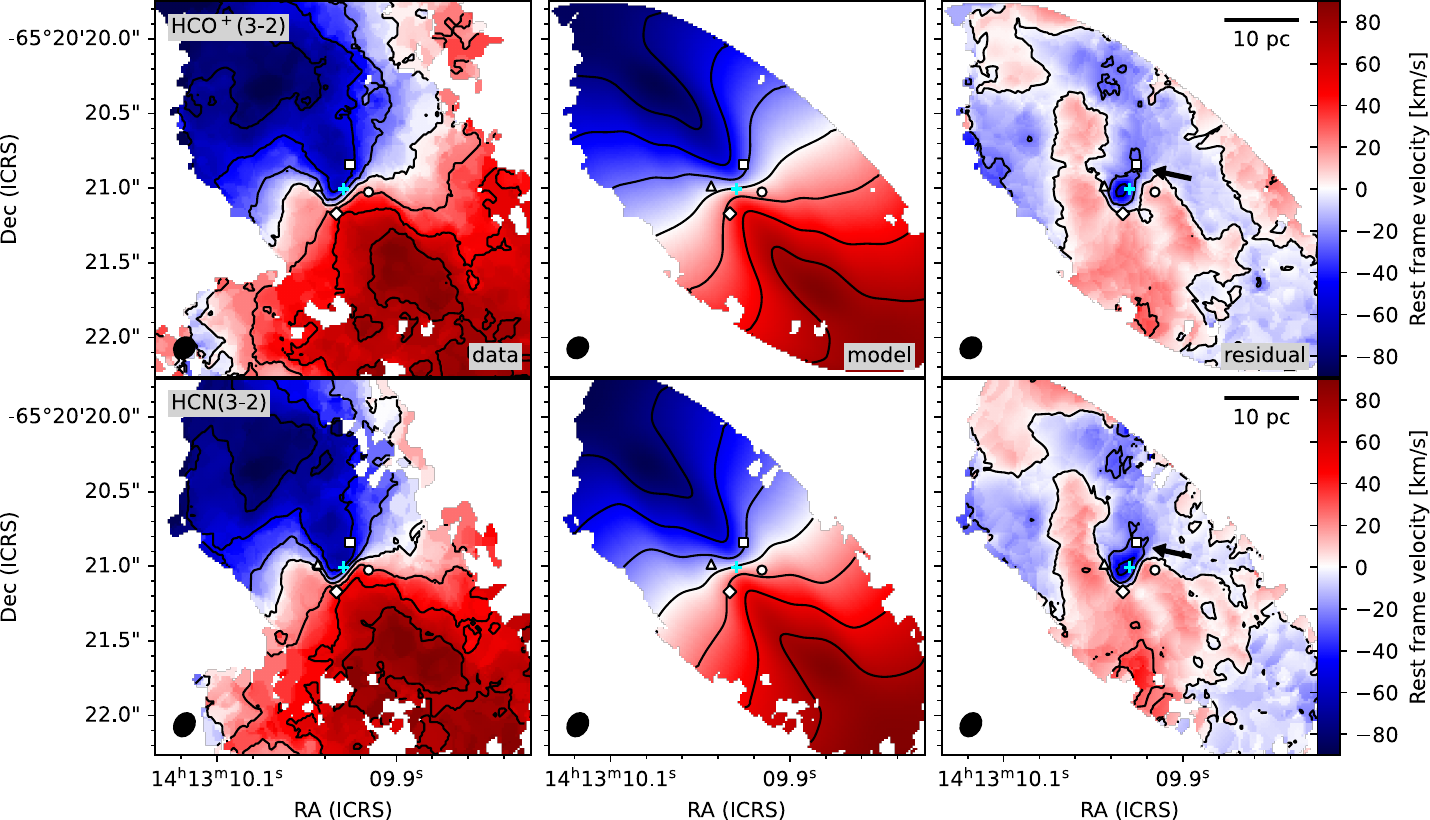}
    \caption{Same as Fig. \ref{fig:BBAROLO_velfield}, but for $V_{\mathrm{rad}}$ left as a free parameter. We find that the velocity residuals become smaller in general, with the blue residual being significantly less pronounced.}
    \label{fig:BBAROLO_velfield_Vradfree}
\end{figure*}

\begin{figure}[h]
    \centering
    \includegraphics[width=\linewidth]{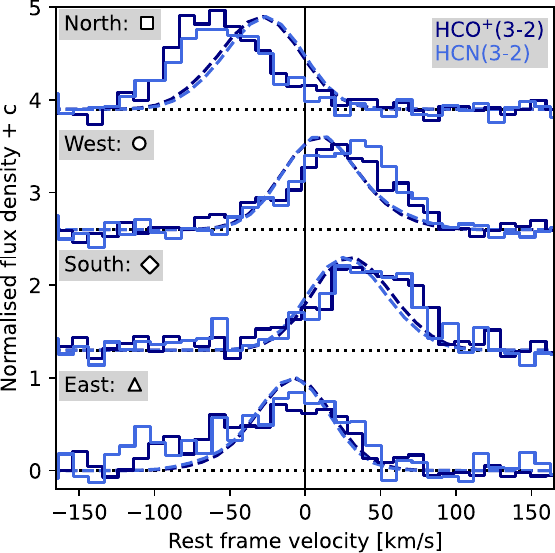}
    \caption{Same as Fig. \ref{fig:BBAROLO_velfield}, but for $V_{\mathrm{rad}}$ left as a free parameter. We find that the model better represents the 1-D kinematic profiles at these four positions but that the model is still unable to produce a sufficiently blue-shifted profile in the northern point aperture.}
    \label{fig:1D_kin_profiles_Vradfree}
\end{figure}

In order to check whether the $^\mathrm{3D}$BAROLO tilted ring fit favours net inflow within the CND when we allow the model to include radial in- or outflow, we perform a second fit. We use the exact same setup as in Sect. \ref{sec:CND_kinematics}, except that we leave $V_{\mathrm{rad}}$ as a free parameter with an initial guess of $V_{\mathrm{rad}}=0$. The resulting radial parameter profiles, the velocity fields and the 1-D kinematic profiles extracted from the four point apertures are presented in Figs. \ref{fig:BBAROLO_params_Vradfree}, \ref{fig:BBAROLO_velfield_Vradfree} and \ref{fig:1D_kin_profiles_Vradfree}, respectively.

\FloatBarrier

We find that the fitted parameters are in general very similar to those for the initial fit without $V_{\mathrm{rad}}$ (i.e., $V_{\mathrm{rad}}=0$), except for the rotation velocities of the inner three rings, which are lower than for $V_{\mathrm{rad}}=0$. Since the projected minor axes of these rings are unresolved, we suspect that the model is not able to properly distinguish between $V_{\mathrm{rot}}$, $V_{\mathrm{disp}}$ and $V_{\mathrm{rad}}$ at this inner region. We therefore do not trust these fitting results for the inner rings\footnote{We note that the inner rings are not relevant for the fitting of the Miyamoto-Nagai model since we fixed the mass of the SMBH, which is the dominant source of gravity at these radii.}. In the bottom panel of Fig. \ref{fig:BBAROLO_params_Vradfree}, we show the radial inflow rate. This was calculated using:

\begin{equation}\label{eq:inflow_radialprofile}
    \begin{aligned}    \dot{M}_{\mathrm{inflow}}(r)=-V_{\mathrm{rad}}(r)\times\frac{d}{dr}\left[M_{\mathrm{MN}}(r)\right],
    \end{aligned}
\end{equation}

where $M_{\mathrm{MN}}(r)$ is the mass enclosed within radius $r$ in the Miyamoto-Nagai model based on the rotation curve from the $^\mathrm{3D}$BAROLO fit with $V_{\mathrm{rad}}=0$.

The residual velocity fields and 1-D kinematic profiles (Figs. \ref{fig:BBAROLO_velfield_Vradfree} and \ref{fig:1D_kin_profiles_Vradfree}) show that the model where $V_{\mathrm{rad}}$ was left free fits the observed kinematics better. We find that the residuals in the velocity field are significantly smaller ($\sim25\,\mathrm{km\,s}^{-1}$ rather than $\sim40\,\mathrm{km\,s}^{-1}$ in the NW-hook), and we find that the 1-D kinematic profiles are better fitted when leaving $V_{\mathrm{rad}}$ free. However, the blue and red residuals along the NW-hook and the E-ridge do not disappear completely, and the northern point aperture still shows a significant discrepancy in Fig. \ref{fig:1D_kin_profiles_Vradfree}. From this, we conclude that the kinematics of the CND is more complex than a simple axisymmetric radial inflow and that the inflow through the spiral structure is faster than the average axisymmetric inflow modelled by $^\mathrm{3D}$BAROLO, meaning that the majority of feeding happens through the spiral arms.

\subsection{Mass inflow as a function of radius}\label{sec:mass_inflow_profile}
The resulting mass inflow profile (bottom panel Fig. \ref{fig:BBAROLO_params_Vradfree}) is relatively flat between a radius of $r=3\,\mathrm{pc}$ and $r=14\,\mathrm{pc}$ at $\dot{M}_{\mathrm{inflow}}=V_{\mathrm{rad}}(r)\times d/dr\left[M_{\mathrm{M-N}}(r)\right]\times f_{\mathrm{gas}}=7.5\pm1.5\,\mathrm{M}_{\odot}\mathrm{yr}^{-1}$. The inner two rings show a significantly lower inflow rate. We again do not consider these values trustworthy since the sizes of their projected minor axes are unresolved. We also observe a drop in the inflow rate beyond $r\approx14\,\mathrm{pc}$. Taken at face value, this could be interpreted as a disruption of the supply of new material outside of this radius or that, assuming steady state, the molecular backflow of the fountain falls back around $ r\sim14\,\mathrm{pc}$, causing the inflow to be larger inside than outside $r\sim14\,\mathrm{pc}$. We need to be cautious, however, since we also observe that the disk structure becomes more complex at this distance (see Figs. \ref{fig:mol_morphology} and \ref{fig:mol_morphology_B7}). Outside of $r\sim14\,\mathrm{pc}$, larger-scale nuclear arms merge with the CND, which we expect to disrupt the dynamics of the disk. In addition, the "SW-ridge" is located around this distance, which contains two distinct velocity components \citepalias{Tristram_2022}. The $^\mathrm{3D}$BAROLO model is not designed to handle such kinematic complexity. Therefore, we do not consider the mass inflow rates at radii larger than $\sim14\,\mathrm{pc}$ trustworthy.

We expect the derived total inflow rates to be an overestimate because the kinematics based on HCO$^{+}$(3-2) and HCN(3-2) traces high-density gas, which is predominantly concentrated in the spiral arms. When multiplying $V_{\mathrm{rad}}$ by the density of the Miyamoto-Nagai model, which contains all material (high- and low-density gas), we effectively assume that all material is falling in at the same velocity as the high-density gas. This is unlikely, since we observe $V_{\mathrm{rad}}$ to be higher in the denser spiral structures (e.g., $V_{\mathrm{rad}}=-43\,\mathrm{km\,s^{-1}}$ in the NW hook) compared to the disk average ($V_{\mathrm{rad}}\sim-12\,\mathrm{km\,s^{-1}}$, Sect. \ref{sec:Vrad_free}). This causes an upward bias in the total inflow rate calculated using Eq. \ref{eq:inflow_radialprofile}.

\end{appendix}

\end{document}